\documentclass[reprint,superscriptaddress,aps,prx,longbibliography]{revtex4-1}

\usepackage{amsmath}
\usepackage{amssymb}
\usepackage{amsthm}
\usepackage{mathtools}
\usepackage{mathdots}
\usepackage{amsfonts}
\usepackage{bbm}
\usepackage{xr}
\usepackage{bbold}
\usepackage{newpxtext}
\usepackage{epsfig,color,dsfont,upgreek,physics}
\usepackage{mathrsfs}
\usepackage{multirow}
\usepackage{graphicx}
\usepackage{dcolumn} 
\usepackage{bm} 
\usepackage{float} 
\usepackage[driverfallback=dvipdfm]{hyperref} 
\usepackage{longtable}
\usepackage{ulem}  
\usepackage{enumerate}
\allowdisplaybreaks

\newcommand{\sgn}{\mathcal{\text{sgn}}}

\renewcommand\[{\begin{equation}}
\renewcommand\]{\end{equation}} 

\def\thefootnote{*}\footnotetext{These authors contributed equally to this work.}

\begin{document}

\title{Josephson Diode Effect from Nonequilibrium Current in a Superconducting Interferometer}

\author{Daniel Shaffer\thefootnote{}}
\affiliation
{Department of Physics, University of Wisconsin-Madison, Madison, Wisconsin 53706, USA}

\author{Songci Li\thefootnote{}}
\affiliation
{Center for Joint Quantum Studies, Department of Physics and Tianjin Key Laboratory of Low-Dimensional Materials Physics and Preparation Technology, Tianjin University, Tianjin 300354, China}

\author{Jaglul Hasan}
\affiliation
{Department of Physics, University of Wisconsin-Madison, Madison, Wisconsin 53706, USA}

\author{Mikhail Titov}
\affiliation{Radboud University, Institute for Molecules and Materials, Heyendaalseweg 135, 6525 AJ Nijmegen, The Netherlands}

\author{Alex Levchenko}
\affiliation
{Department of Physics, University of Wisconsin-Madison, Madison, Wisconsin 53706, USA}

\begin{abstract}
We investigate the Josephson diode effect in a superconducting interferometer under nonequilibrium conditions. In contrast to its thermodynamic counterpart, which requires the simultaneous breaking of time-reversal and inversion symmetry, we demonstrate that a diode-like asymmetry of the critical current can emerge solely due to a dissipative current in the normal region of an otherwise symmetric Josephson junction. This effect is driven entirely by the nonequilibrium conditions, without the need for additional inversion symmetry breaking. Using the standard quasiclassical Keldysh Green’s function formalism, we explicitly calculate the diode coefficient from the supercurrent-phase relation of the interferometer. Remarkably, within certain ranges of control parameters, such as applied voltage, temperature, and the geometric aspect ratio of the device, the diode coefficient can exceed its nominal perfect value.
\end{abstract}
\date{August 20, 2025}
\maketitle


\section{Introduction}

Many different mechanisms for realizing superconducting and Josephson diode effects (SDE and JDE; we henceforth reserve the former term for bulk superconductors) have been proposed in recent years
~\cite{Nadeem23, NagaosaYanase24, Ma25, LevitovNazarovEliashberg85, Edelstein96, YuanFu22, IlicBergeret22, DavydovaFu22, HasanSongciLevchenko22, BanerjeeScheurer24alt, HasanShafferKhodasLevchenko24, ChakrabortyBlackSchaffer24, IlicBergeret24, Ghosh24, ShafferChichinadzeLevchenko24, HasanShafferKhodasLevchenko25, BankierLevchenkoKhodas25}. Defined by the inequality in the magnitudes of the critical supercurrents \(J_{c+}\) and \(J_{c-}\) flowing in opposite directions, these effects require the simultaneous breaking of time-reversal and inversion symmetries. In the overwhelming majority of experimental realizations and theoretical proposals, time-reversal symmetry is broken by either external or, less commonly, internal magnetic fields~\cite{Nadeem23, Ma25}.
Inversion symmetry is typically broken by spin-orbit coupling (SOC) or, in the case of JDE, by junction geometry asymmetries~\cite{Krasnov97, ZinklSigrist22}. Other extrinsic effects, such as vortex motion, have also been invoked~\cite{Goldman67, LeeBarabasi99, GutfreundBuzdin23, MollGeshkenbein23}.

An alternative route to breaking both time-reversal and inversion symmetry involves applying an external electric current, either a normal or a supercurrent. For instance, an external supercurrent has been identified as the source of JDE in field-free multiterminal Josephson junctions recently realized in several experiments~\cite{GuptaPribiag23, ChilesFinkelstein23, Zhang24}. This mechanism was explicitly emphasized in~\cite{ChilesFinkelstein23}, where it was demonstrated that an external supercurrent can control the JDE in an otherwise symmetric three-terminal device, achieving a diode with \emph{perfect} efficiency, \(\eta = (J_{c+} + J_{c-}) / (J_{c+} - J_{c-}) = \pm 1\), so that the supercurrent flows only in one direction. JDE realizations in SQUID devices can also be qualitatively attributed to circulating supercurrents~\cite{DeWaele67, RaissiNordman94, CiacciaManfra23, ReinhardtManfra24, CuozzoShabani24}; however, in those cases explicit symmetry breaking by magnetic fields and device asymmetry is still required. In theoretical works \cite{DavydovaFu22, DavydovaFu24} JDE was attributed to Meissner screening supercurrents and modeled via finite-momentum Cooper pairs. Several proposals involving supercurrent feedback, effectively using three-terminal devices ~\cite{GuarcelloFilatrella24, MarginedaGiazotto25}, have been made. Additionally, a realization of the perfect diode effect due to a persistent supercurrent in chiral nanotube has also been proposed \cite{Cuozzo25}.

The effects of external normal currents have also been considered. Curiously, this includes what may be the first experimental report of a superconducting diode effect, where ``pinch'' currents through normal metal wires adjacent to a thin-film superconductor induced diode behavior~\cite{EdwardsNewhouse62}. However, in that case the effect can also be attributed to magnetic fields generated by the currents.

A major theoretical challenge in treating external currents is the necessity of accounting for nonequilibrium phenomena. To date, only one theoretical study has explicitly connected a dissipative current to a nonreciprocal parallel supercurrent, using a time-dependent Ginzburg-Landau (TDGL) approach~\cite{BanerjeeScheurer25}. An earlier study investigated SDE in a bilayer system with a perpendicular electric field driving a supercurrent between layers~\cite{DaidoYanase25} (symmetry-wise, equivalent to Rashba SOC). Both studies employed phenomenological TDGL models and required additional inversion symmetry breaking, either through valley polarization~\cite{BanerjeeScheurer25} or other mechanisms~\cite{DaidoYanase25}, and both predicted the possibility of \emph{perfect} diode efficiency (also referred to as unidirectional superconductivity in~\cite{DaidoYanase25}).

Nonequilibrium effects have otherwise primarily been explored in the context of AC-driven Josephson junctions. In that case, an AC-version of JDE can be defined as the nonreciprocity of the DC (time-averaged) voltage response under bias current. AC JDE has been observed in several experiments~\cite{MendittoGoldobin16, SchmidGoldobin24, Su24, MatsuoManfra25, BorgonginoGiazotto25} and studied theoretically, mostly through phenomenological RCSJ models or numerical Bogoliubov–de Gennes (BdG) simulations~\cite{SeoaneSoutoDanon24, BorgonginoGiazotto25, MonroeZutic24, Soori23I}, with some analytical treatments also available~\cite{OrtegaTaberner23, LiuAndreevSpivak24, ZazunovEgger24}. Related phenomena have been investigated in multiterminal devices and SQUIDs~\cite{Zapata96, Goldobin01, Carapella01, Lee03, GolodKrasnov22, SeoaneSouto22, SeoaneSoutoDanon24, GuarcelloFilatrella24, ValentiniDanon24, CuozzoShabani24}. In this context, regimes with \(|\eta| = 1\) and even \(|\eta| > 1\) have been both theoretically predicted~\cite{Zapata96, SeoaneSoutoDanon24, BorgonginoGiazotto25} and experimentally observed~\cite{ValentiniDanon24, Su24, BorgonginoGiazotto25}. However, as this effect is dissipative and pertains only to time-averaged quantities, the relevance of AC JDE for equilibrium superconducting diode applications remains limited.
Finally, asymmetric dissipative currents can give rise to the non-reciprocity of retrapping currents in overdamped Josephson junctions even without time reversal symmetry breaking \cite{TrahmsVonOppen23, SteinerVonOppen23}, but the critical currents remain reciprocal in that case and no JDE occurs in the strict sense.

\begin{figure}[t]
\includegraphics[width=0.48\textwidth]{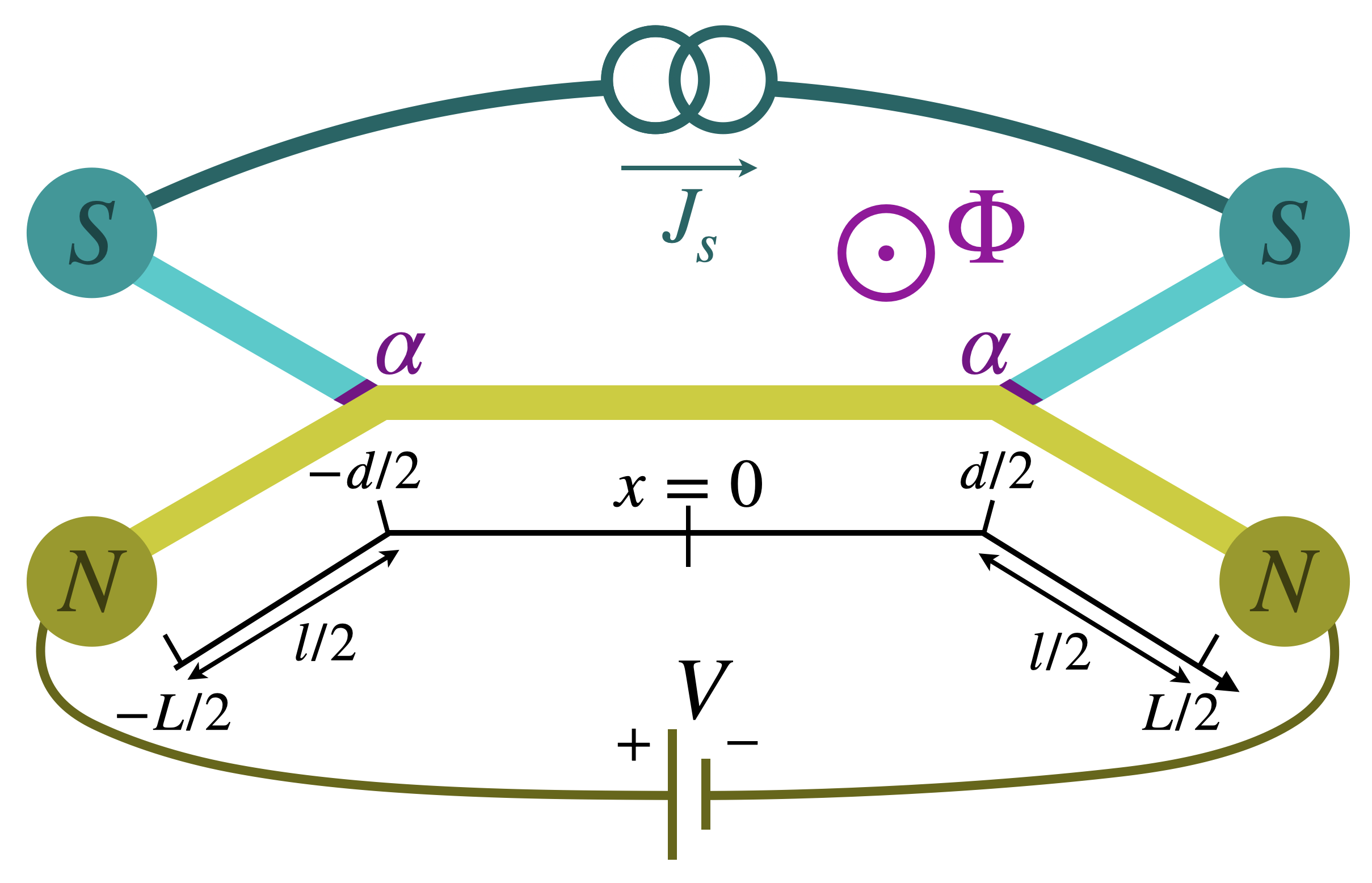} 
\caption{Schematic of the Andreev interferometer considered in this work. Blue and brown circles denote superconducting and normal-metal reservoirs, respectively, acting as device terminals. A normal-metal wire of length \(L\) connects the two normal terminals, while the superconducting reservoirs are coupled via low-transparency NS interfaces with transparency \(\alpha\), forming an SNS junction. A DC bias \(V\) is applied between the normal terminals at a distance \(l/2\) from the NS interface. A supercurrent \(J_s\) or, equivalently, a phase difference \(\varphi\) --- controlled by the external flux \(\Phi\) threading the interferometer --- is imposed between the superconducting terminals.}  \label{fig1}
\end{figure}

In this work, we revisit a previous theoretical calculation to demonstrate explicitly that the conventional (DC) Josephson diode effect (JDE) can arise from a dissipative current in the normal region of an otherwise symmetric Josephson junction, with the effect being \emph{entirely} driven by nonequilibrium conditions and requiring no additional symmetry breaking. To the best of our knowledge, this possibility has not been previously noted in the literature.

We consider a model describing a standard Andreev interferometer, consisting of a normal-metal wire in the diffusive (i.e., strongly disordered) regime connecting two normal-metal reservoirs, with two low-transparency NS interfaces symmetrically attached to two superconducting reservoirs along the wire (see Fig.~\ref{fig1}). This system has been extensively studied in the past by several authors~\cite{Volkov95, GolubovWilhelmZaikin97, WilhelmZaikin98, VirtanenHeikkila04, Titov08, DolgirevKalenkovZaikin18, KalenkovZaikin21}. However, since those works predate the recent surge of interest in superconducting diodes and primarily focused on thermoelectric effects, the presence of the JDE in these devices was not recognized, despite being implicitly contained in their results. Only the related anomalous Josephson effect (AJE) --- characterized by an asymmetry in the current-phase relation (CPR), with \(J(\varphi)\) vanishing at \(\varphi = \varphi_0 \neq 0, \pi\) --- was explicitly reported in~\cite{DolgirevKalenkovZaikin18}. Variants of Andreev interferometers with different configurations have also been studied~\cite{WilhelmZaikin98, Baselmans99, ShaikhaidarovVolkov00, Bezuglyi03, SunLinder24}, and AJE has been noted in some of them~\cite{HijanoIlicBergeret21, MarginedaCheckley23}.

We explicitly compute the CPR of the interferometer using the standard quasiclassical nonequilibrium Keldysh Green's function formalism, closely following the approach of~\cite{Titov08}. From the CPR, we identify the maximum \(J_{c+}\) and minimum \(J_{c-}\) supercurrents and evaluate the diode efficiency \(\eta\). Remarkably, we not only find a finite diode effect \((|\eta| > 0)\), as expected from symmetry arguments, but also observe regimes of perfect \((|\eta| = 1)\) and even ``supra-perfect'' \((|\eta| > 1)\) diodicity, depending on system parameters such as applied voltage, temperature, and the distance between the interfaces and the terminals. In particular, one of the critical currents \(J_{c\pm}\) first vanishes and then changes sign, resulting in a regime where strictly nonzero supercurrents are allowed only in a single direction. Unlike the case of AC JDE, here the voltage between the superconducting terminals remains exactly zero between the critical currents at all times, rather than only on average. Interestingly, a similar phenomenon has been reported in bulk superconducting diode effect measurements in twisted trilayer graphene~\cite{DaidoYanaseLaw25, LinScheurerLi22}.


\section{Usadel Equation and Boundary Conditions}

We model the Andreev interferometer using the standard quasiclassical nonequilibrium Green's function \(\check{G}(x,\varepsilon)\), with \(x\in(-L/2,L/2)\) denoting the coordinate along the diffusive normal metal wire; we follow closely the formalism developed in \cite{Titov08}. Within the wire, the Green's function satisfies the Usadel equation (here we neglect inelastic processes):
\begin{equation}
\label{Usadel}
D\frac{d \check{I}}{dx}+[i \varepsilon \tau_z, \check{G}]=0,
\quad \check{I}\equiv \check{G}\frac{d \check{G}}{dx},
\end{equation}
where $D$ is the diffusion coefficient and $\check{I}$ is the matrix current; \(\tau_z\) is the Pauli \(z\) matrix in Nambu space. The nonequilibrium Green's function is a matrix in Keldysh space and satisfies the normalization condition:
\begin{equation}
\label{quasi}
\check{G}=\left(
\begin{array}{cc}\hat{G}^R& \hat{G}^K\\ 0& \hat{G}^A \end{array}
\right), 
\qquad \check{G}^2=1.
\end{equation}
In addition, to describe the proximity effect, \(\hat{G}^{R/A/K}\) are matrices in Nambu space:
\begin{equation}
\label{prmRA}
\hat{G}^{R (A)} = \left(\begin{array}{cc} \pm g^{R (A)} & f^{R (A)} \\ 
\bar{f}^{R (A)} & \mp g^{R (A)} \end{array}\right), 
\end{equation}
where the bar denotes charge conjugation: \(\bar{f}^{R (A)}(x,\varepsilon)=(f^{R (A)}(x,-\varepsilon))^*\). The anomalous components \(f^{R (A)}\) are nonzero in the presence of superconducting correlations. The Keldysh component of the normalization constraint reads  \(\hat{G}^R \hat{G}^K+ \hat{G}^K \hat{G}^A =0\) and implies that \(\hat{G}^K\) has only two linearly independent entities \cite{Kopnin01}: in the standard parameterization 
\(\hat{G}^K=\hat{G}^R \hat{h}-\hat{h}\hat{G}^A\) with the diagonal matrix 
\begin{equation}
\hat{h}=h+\tau_z h_\sigma, 
\end{equation} 
where \(h\)  and \(h_\sigma\) are the two independent components of the distribution function.

The boundary conditions are given by the Green's functions within the normal and superconducting reservoirs, assumed to be in equilibrium. The two normal reservoirs at \(x=\mp L/2\) are thus described by \(\hat{G}^{R(A)}=\pm \tau^z\) and distribution functions
\begin{subequations}
\[\label{h_conditions}
\left.\hat{h}\right|_{x\to -L/2,L/2}=
\left(\begin{array}{cc} h_{1,2} & 0\\ 0 & \bar{h}_{1,2} \end{array} \right)\,,\]
where
\begin{equation}
h_a=\tanh\frac{\varepsilon-\mu_a}{2T_a}, \qquad
\bar{h}_a=\tanh\frac{\varepsilon+\mu_a}{2T_a},
\end{equation}
\end{subequations}
with chemical potentials $\mu_1$, $\mu_2$ and temperatures $T_1$, $T_2$, and $a=1,2$. In particular, the functions $(1-h_a)/2$, $(1-\bar{h}_a)/2$ 
are Fermi distribution functions of electrons and holes, 
respectively.

For the superconducting reservoirs, we assume that the SC gap \(\Delta\) drops abruptly to zero at the NS interfaces, neglecting small corrections that would come from imposing self-consistency. Working only at temperatures much lower than the SC gap and assuming the latter is much larger than the Thouless energy \(E_T=D/L^2\), i.e., \(\{T, E_T\} \ll \Delta\), we can take
\[\label{gS}
\hat{G}^{R}_S=\hat{G}^{A}_S=\left(\begin{array}{cc} 0 & e^{i\chi}\\ e^{-i\chi}& 0\end{array}\right),\]
where \(\chi\) is the phase of the SC order parameter. Taking the chemical potential of the Cooper pairs to be zero without loss of generality and neglecting quasiparticle effects, we can moreover take \(\hat{G}^{K}_S=0\) (see \cite{Titov08} for details). The difference between the phases \(\chi_1\) and \(\chi_2\) at the left and right NS interfaces can be fixed by an external magnetic flux \(\Phi\): \(\varphi=\chi_1-\chi_2=2\pi \Phi/\Phi_0\), where \(\Phi_0=h/(2e)\) is the flux quantum. Alternatively, in a supercurrent biased setup the phase is determined self-consistently by solving for the current as a function of \(\varphi\).

In addition, the reservoir boundary conditions must be supplemented by boundary conditions at the NS interfaces. In particular, we assume that the interfaces have low transparency as parameterized by a coefficient \(\alpha=T_B/l_B\), where \(T_B\) is the transmission probability of the barrier and \(l_B\) is its effective length (\(\alpha\) thus has dimensions of inverse length). This allows us to impose the rigid Kupriyanov-Lukichev 
boundary conditions \cite{GolubovKupriyanov04}: \(\check{G}\) is continuous in \(x\) inside the normal wire, but its derivative can be discontinuous at the position of the NS interfaces as \(x=\pm d/2\); more precisely, the matrix current satisfies
\begin{align}
    \delta[\check{I}]_{\pm d/2}&\equiv\lim_{\delta\rightarrow0}\left[\check{I}(\pm d/2+\delta)-\check{I}(\pm d/2-\delta)\right]=\nonumber\\
    &=\frac{\alpha}{2}[\check{G}_S,\check{G}]\,,
\end{align}
where \(\check{G}_S\) is given by Eq. \ref{gS}. The latter condition imposes current conservation at the interface. Furthermore, we assume charge conservation in the superconducting arm of the interferometer, which implies that the current coming in at one interface must be equal to that coming out at the other: \(\int (\delta[j^e(\varepsilon)]_{d/2}+\delta[j^e(\varepsilon)]_{-d/2})d\varepsilon=0\), where the energy-resolved electric current density is given by
\[j^e(\varepsilon)=\Tr[\tau^z\hat{I}^K]\]
with \(\hat{I}^K=\hat{G}^R \partial_x\hat{G}^K + \hat{G}^K \partial_x\hat{G}^A\). Note that here we relax the assumption of pure elastic scattering (see \cite{Titov08} for further discussion). The total current through the normal wire is given by \(J^e=\sigma/(8e)\int j^e(\varepsilon) d\varepsilon\) with conductivity \(\sigma=2e^2 D\nu\) (\(\nu\) being the density of states of the normal metal).

Finally, the weak proximity effect due to the low transparency assumption allows the Usadel equation to be linearized in the anomalous Green's function components \(f^{R(A)}\), which results in
\[\label{UsadelfR}
\frac{d^2f^{R}}{dx^2}=z^2 f^{R}, \qquad z^2\equiv-2i\frac{\varepsilon+i0}{D}\,,\]
(\(i0\) is retained for keeping track of the analytic properties of the Green's function; the advanced component is obtained via \(f^A(\varepsilon)=f^R(-\varepsilon)\)). For purposes of computing the supercurrent, it is sufficient to consider the values of \(f^R\) near the NS interfaces. 
The analytical solution for the boundary conditions specified above takes the form 
\begin{subequations}
\begin{equation}
f^R(\pm d/2)=\frac{L}{\alpha}\left[e^{i\chi_{1/2}}F_1+e^{i\chi_{2/1}}F_2\right],
\end{equation}
where
\begin{align}
    F_1&=\alpha^2\frac{\sinh(zl/2) \sinh[z(L-l/2)]}{zL \sinh(zL)},\\
    F_2&=\alpha^2\frac{\sinh^2(zl/2)}{zL \sinh(zL)}.
\end{align}
\end{subequations}
Within the same low transparency approximation, it is sufficient to compute \(\hat{h}\) in the normal wire in the absence of the proximity effect, given by:
\begin{subequations}
\begin{align}
    h(\varepsilon,x)=\frac{1}{4}\left(h_0-\frac{2x}{L}h_T\right),\\
    h_\sigma(\varepsilon,x)=\frac{1}{4}\left(h_\mu-\frac{2x}{L}h_{\mu T}\right),
\end{align}
\end{subequations}
where \(h_0=\sum_{a}(h_a+\bar{h}_a)\), \(h_T=-\sum_{a}(-1)^a(h_a+\bar{h}_a)\), \(h_\mu=\sum_{a}(h_a-\bar{h}_a)\), and \(h_{\mu T}=-\sum_{a}(-1)^a(h_a-\bar{h}_a)\). In the absence of a thermal gradient, i.e., when \(T_1=T_2\), \(h_T=0\) and the charge conservation boundary condition implies that \(h_\mu=0\) as well. It is, however, necessary to compute the leading corrections to the discontinuity of the derivatives of the distribution functions at the NS interfaces:
\[\delta\left[\frac{dh_\sigma}{dx}\right]_{\pm \frac{d}{2}}=L h_\sigma(\pm d/2)\Re\left[F_1+F_2\cos\varphi\right].\]
Note that since \(h_\sigma\) encodes the electron/hole asymmetry of the distribution function, the dissipative part of the electric current is precisely proportional to its spatial derivative \(\frac{dh_\sigma}{dx}\). The above discontinuities therefore encode the portion of the electric current flowing from the normal wire into the superconductor at the interface and, as we will see below, give rise to the nonequilibrium corrections to the supercurrent.

\begin{figure*}[t!]
\includegraphics[width=0.99\textwidth]{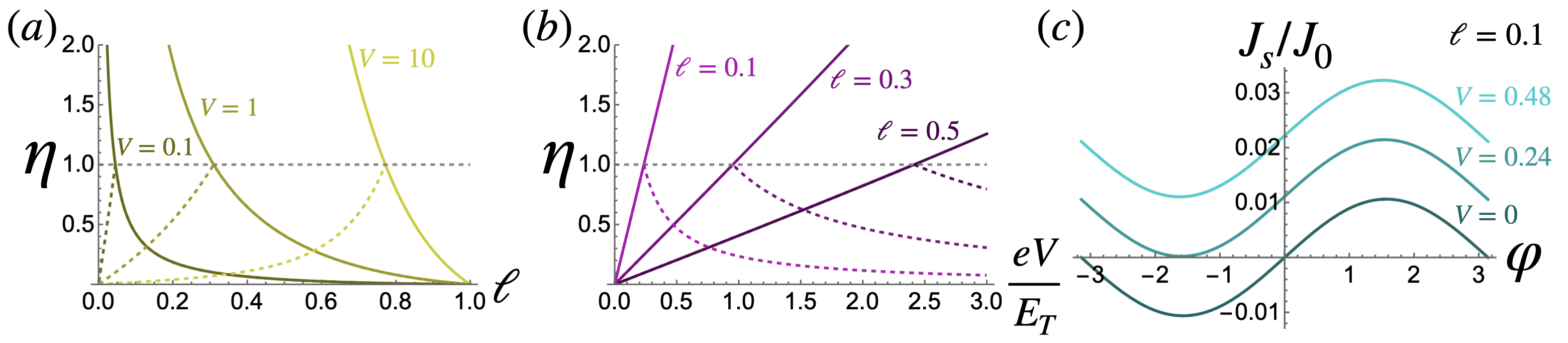} 
\caption{(a-b) Plots of the Josephson diode efficiency coefficient \(\eta\) defined in Eq. (\ref{eta}) versus \(\ell=l/L\) (\(V\)) for various values of \(V\) (\(\ell\)); voltage is in units of \(E_T/e\), \(T=0.001E_T\), and note that \(\eta\) is independent of \(\alpha\). The perfect diode effect corresponds to \(\eta=1\), shown as a gray dashed line. Dashed curves show the diode efficiency coefficient \(\eta'\) with the alternative definition Eq. (\ref{etaprime}). (c) Representative CPR curves at \(l/L=0.1\) without any diode effect (\(eV=0\)), with the perfect Josephson diode effect (\(eV=0.24 E_T\)) at which \(J_{c-}=0\), and the supra-perfect diode effect (\(eV=0.48E_T\)). The current density is measured in units of \(J_0=\frac{G_NE_T}{eS} (\alpha L)^2\) (see Eq. (\ref{EqJ})).}  \label{fig2}
\end{figure*}

\section{Supercurrent and the Josephson Diode Effect out of Equilibrium}

The supercurrent flowing in the Andreev interferometer as a function of phase bias has been calculated to second order in the transmission coefficients \(\alpha\) in \cite{Titov08}, and for the symmetric geometry with \(\mu_1=-\mu_2=eV/2\) and no thermal bias. It can be decomposed into equilibrium and nonequilibrium parts as follows:
\[J_s(\varphi)=J_{s}^{(eq)}(\varphi)+J_{s}^{(neq)}(\varphi)\]
where \(J_{s}^{(eq)}(\varphi)=J_{s,1}^{(eq)}\sin\varphi\). The nonequilibrium part comes from corrections to the distribution functions, in particular the discontinuity in their derivatives due to the proximity effect at the NS interfaces:
\begin{align}
    J_s^{(neq)}(\varphi)&=\frac{\sigma}{4e}\int d\varepsilon\left[\left(\delta\left[\frac{dh_\sigma}{dx}\right]_{\frac{d}{2}}-\delta\left[\frac{dh_\sigma}{dx}\right]_{-\frac{d}{2}}\right)\right]\nonumber\\
    &=J_{s,0}^{(neq)}+J_{s,2}^{(neq)}\cos\varphi\,.
\end{align}
The coefficients can be computed as integrals of the distribution functions and anomalous parts of the quasiclassical Green's function:
\begin{subequations}
\begin{align}
    J_{s,1}^{(eq)}&=\frac{\sigma L}{8e}\int  h_0\Im F_2 d\varepsilon \\
    J_{s,0}^{(neq)}&=-\frac{\sigma d}{8e}\int  h_{\mu T}\Re F_1 d\varepsilon \\ 
    J_{s,2}^{(neq)}&=-\frac{\sigma d}{8e}\int  h_{\mu T}\Re F_2 d\varepsilon
\end{align}
\end{subequations}
Importantly, both nonequilibrium terms are anomalous: the cosine term \(J_{s,2}^{(neq)}\) results in the anomalous Josephson effect with a phase shift \(\varphi_0=\tan^{-1}(J_{s,2}^{(neq)}/J_{s,1}^{(eq)})\), as noted earlier in \cite{DolgirevKalenkovZaikin18}, while the constant term \(J_{s,0}^{(neq)}\) results in the Josephson diode effect. In particular, the diode efficiency coefficient is given by
\[\eta=\frac{J_{c+}+J_{c-}}{J_{c+}-J_{c-}}=\frac{J_{s,0}^{(neq)}}{\sqrt{J_{s,1}^{(eq)2}+J_{s,2}^{(neq)2}}}\label{eta}\]
with \(J_{c\pm}\) being the maximum and the minimum of \(J_s(\varphi)\), respectively.

The integrals can in principle be evaluated analytically in terms of the hypergeometric functions. In the limit of low temperature and voltage (compared to \(E_T\)), we find
\begin{subequations}\label{EqJ}
\begin{align}
J_{s,1}^{(eq)}&\approx\frac{J_0}{2\sqrt{2}}\left[\psi_+\left(\frac{\ell}{2}\right)+\psi_-\left(\frac{\ell}{2}\right)-2\psi\left(\frac{1}{2}\right)\right]\\
J_{s,0}^{(neq)}&\approx\frac{J_0}{2}\frac{\ell(1-\ell)(2-\ell)}{4}\frac{eV}{E_T}\\
J_{s,2}^{(neq)}&\approx\frac{J_0}{2}\frac{\ell^2(1-\ell)}{4}\frac{eV}{E_T}
\end{align}
\end{subequations}
where \(\ell=l/L\) characterizes the interferometer geometry, \(\psi\) is the digamma function and we used the notation $\psi_\pm(x)=\psi\left(1/2\pm x\right)$; for clarity, we also defined \(J_0=\frac{G_NE_T}{eS} (\alpha L)^2\), where \(S\) is the cross-sectional area of the normal metal wire, such that \(G_N=S\sigma/L\) is its conductance. It is worth noting that the dimensionless product $\alpha L$ in the current density $J_0$ is not necessarily small even in the tunneling limit $T_B\ll1$. Indeed, the interfacial parameter can be expressed as a ratio of the junction areal conductance to the conductivity of a superconductor in the normal state, namely $\alpha=g/S\sigma$. The former can be estimated from the Landauer formula $g\sim e^2 T_B N_{ch}$, with the number of channels in the junction $N_{ch}\sim Sk^2_F$, whereas the latter can be estimated from the Drude formula $\sigma\sim e^2\nu D$. Therefore, $\alpha L=gL/S\sigma\sim T_B (L/l_{tr})$, where $l_{tr}$ is the transport mean free path, which is short in the diffusive limit so that $L/l_{tr}\gg1$. 

Importantly, for \(\ell\rightarrow 0\) (i.e., in the long junction limit), \(J_{s,0}^{(neq)}\propto \ell\), while \(J_{s,1}^{(eq)},J_{s,2}^{(neq)}\propto \ell^2\). As a result, \(\eta\rightarrow \infty\) as \(\ell\rightarrow 0\), resulting in the perfect JDE for some intermediate \(\ell\) and supra-perfect JDE below it, for \emph{any} value of the fixed voltage. We confirm numerically that this qualitative behavior persists for arbitrary temperatures and voltages below the gap, as shown in Fig. \ref{fig2} (a). 
 As a function of \(V\), \(\eta\) is approximately linear, as seen in Fig. \ref{fig2} (b). This is reflected in the CPR being shifted vertically by an amount proportional to \(V\), as shown in Fig. \ref{fig2} (c) (there is also a smaller horizontal shift due to \(J_{s,2}^{(neq)}\)). Consequently, at some value of \(V)\), the global minimum \(J_{c,-}\) crosses zero, above which a nondissipative supercurrent is only possible in one direction and the supra-perfect JDE is realized.
For this supra-perfect JDE, it can be more useful to redefine the diode efficiency as
\[\eta'=\frac{|J_{c+}|-|J_{c-}|}{|J_{c+}|+|J_{c-}|}=\eta^{\sgn[1-|\eta|]}\label{etaprime}\]
This agrees with \(\eta\) when \(|\eta|\leq1\). When \(|\eta|>1\), \(\eta'\) better captures the effectiveness of the Josephson diode, as it vanishes if \(J_{c+}=J_{c-}\), in which case \(J_s\) is independent of \(\varphi\) and is no longer useful for diode applications. In particular, \(\eta'\rightarrow 0\) as \(l/L\rightarrow0\) in our case (see Fig. \ref{fig2}(a-b)).

\section{Summary and Discussion}

The general result \(J_s = J_{s,0}^{(neq)} + J_{s,1}^{(eq)} \sin\varphi + J_{s,2}^{(neq)} \cos\varphi\) can be understood in terms of the underlying physical processes contributing to each term. The term \(J_{s,1}^{(eq)}\) corresponds to the conventional Josephson effect and arises from Cooper pairs that split into unpaired electrons at one NS interface and recombine at the second. The other two terms originate from Andreev reflection processes, in which an electron is reflected as a hole at one interface, and a hole is reflected as an electron at the other. These reflections can be either correlated or uncorrelated, depending on whether the same quasiparticle participates in both events. 

In the case of correlated (crossed) Andreev reflections, the hole acquires an Aharonov-Bohm phase due to the phase difference \(\varphi\) between the two superconducting reservoirs, leading to the phase-dependent \(J_{s,2}^{(neq)}\) term, as discussed in~\cite{DolgirevKalenkovZaikin18}. By contrast, uncorrelated Andreev reflections generate the phase-independent \(J_{s,0}^{(neq)}\) term, which is responsible for the diode effect~\footnote{From a phenomenological Ginzburg-Landau perspective, the term \(J_{s,0}^{(neq)}\) can be interpreted as arising from a contribution \(\mathcal{F}_{s,0} = J_{s,0}^{(neq)} \varphi\) to the free energy, analogous to the \(\mathbf{J} \cdot \mathbf{A}\) coupling via gauge invariance. This interpretation also explains why no JDE arises solely from a thermal gradient: a heat current \(J_q\) does not couple to the vector potential or, consequently, to the phase difference \(\varphi\). Similar constant terms in the CPR have also been noted to arise due to screening and persistent supercurrents \cite{CuozzoShabani24,Cuozzo25}.}.

In the long-junction limit, processes requiring coherent correlations between the two NS interfaces (namely, Josephson and crossed Andreev reflections) are suppressed relative to uncorrelated Andreev reflections, resulting in an enhanced diode efficiency. Importantly, the Andreev reflections generate a net current only due to the electron-hole imbalance encoded in the nonequilibrium distribution function \(h_\sigma\), confirming that the JDE is a purely nonequilibrium phenomenon.

We further note that the ability to tune the DC JDE via an external voltage applied to the normal region enables the realization of a switchable diode, effectively functioning as a four-terminal superconducting transistor. A potential limitation of such a device is dissipation associated with the normal current; however, this tradeoff may be acceptable in practice. Notably, the voltage required to achieve supra-perfect JDE can be engineered through junction geometry and becomes vanishingly small in the long-junction limit.

In principle, the normal-metal wire could be replaced by a superconducting wire, and the NS interfaces by SIS junctions, resulting in a dissipation-free variant of the device. This setup underlies the operation of four-terminal Josephson junction diodes~\cite{Amin01, PankratovaShabani20, GolodKrasnov22}. In the specific geometry considered here, the four-terminal device would correspond to an SIS'IS junction~\cite{OsinLevchenkoKhodas24}, where the S' region denotes the superconducting wire replacing the normal wire in Fig.~\ref{fig1}.

In our analysis, we have neglected self-field effects arising from current-generated magnetic flux, which could further contribute to diode behavior in interesting ways, as is known to occur in other systems \cite{Krasnov97, VarvaStrunk13, CuozzoShabani24}. We have also neglected possible geometrical effects, which can be taken into account by generalizing the Usadel equations to curved quantum wires \cite{Salamone21,Salamone22}. A detailed exploration of these effects is left for future work.

Finally, we comment on the possible connection between the supra-perfect JDE demonstrated here and the recently observed supra-perfect bulk SDE in twisted trilayer graphene (TTG)~\cite{LinScheurerLi22, DaidoYanaseLaw25}. The perfect SDE in TTG has been attributed to the coupling between valley polarization order and a dissipative current~\cite{BanerjeeScheurer25}, a non-equilibrium mechanism closely analogous to the supra-perfect JDE model discussed here. The supra-perfect SDE has also been modeled via the coexistence of three superconducting states with different Cooper pair momenta in equilibrium~\cite{DaidoYanaseLaw25}. If non-equilibrium effects are nevertheless at play in the observed supra-perfect bulk superconducting diode effects in TTG,
our interpretation of the supra-perfect JDE may offer insight into its underlying mechanism.

\section*{Acknowledgments}

We thank Pavel Dolgirev and Gleb Finkelstein for useful discussions regarding Refs. \cite{DolgirevKalenkovZaikin18} and \cite{ChilesFinkelstein23} respectively. We are also grateful to Joseph J. Cuozzo, Felix von Oppen, and Anatoly F. Volkov for valuable comments. This work was supported by the National Science Foundation Grant No. DMR-2452658 and H. I. Romnes Faculty Fellowship provided by the University of Wisconsin-Madison Office of the Vice Chancellor for Research and Graduate Education with funding from the Wisconsin Alumni Research Foundation.

\bibliography{JDE-biblio}

\begin{thebibliography}{82}%
\makeatletter
\providecommand \@ifxundefined [1]{%
 \@ifx{#1\undefined}
}%
\providecommand \@ifnum [1]{%
 \ifnum #1\expandafter \@firstoftwo
 \else \expandafter \@secondoftwo
 \fi
}%
\providecommand \@ifx [1]{%
 \ifx #1\expandafter \@firstoftwo
 \else \expandafter \@secondoftwo
 \fi
}%
\providecommand \natexlab [1]{#1}%
\providecommand \enquote  [1]{``#1''}%
\providecommand \bibnamefont  [1]{#1}%
\providecommand \bibfnamefont [1]{#1}%
\providecommand \citenamefont [1]{#1}%
\providecommand \href@noop [0]{\@secondoftwo}%
\providecommand \href [0]{\begingroup \@sanitize@url \@href}%
\providecommand \@href[1]{\@@startlink{#1}\@@href}%
\providecommand \@@href[1]{\endgroup#1\@@endlink}%
\providecommand \@sanitize@url [0]{\catcode `\\12\catcode `\$12\catcode
  `\&12\catcode `\#12\catcode `\^12\catcode `\_12\catcode `\%12\relax}%
\providecommand \@@startlink[1]{}%
\providecommand \@@endlink[0]{}%
\providecommand \url  [0]{\begingroup\@sanitize@url \@url }%
\providecommand \@url [1]{\endgroup\@href {#1}{\urlprefix }}%
\providecommand \urlprefix  [0]{URL }%
\providecommand \Eprint [0]{\href }%
\providecommand \doibase [0]{http://dx.doi.org/}%
\providecommand \selectlanguage [0]{\@gobble}%
\providecommand \bibinfo  [0]{\@secondoftwo}%
\providecommand \bibfield  [0]{\@secondoftwo}%
\providecommand \translation [1]{[#1]}%
\providecommand \BibitemOpen [0]{}%
\providecommand \bibitemStop [0]{}%
\providecommand \bibitemNoStop [0]{.\EOS\space}%
\providecommand \EOS [0]{\spacefactor3000\relax}%
\providecommand \BibitemShut  [1]{\csname bibitem#1\endcsname}%
\let\auto@bib@innerbib\@empty
\bibitem [{\citenamefont {Nadeem}\ \emph {et~al.}(2023)\citenamefont {Nadeem},
  \citenamefont {Fuhrer},\ and\ \citenamefont {Wang}}]{Nadeem23}%
  \BibitemOpen
  \bibfield  {author} {\bibinfo {author} {\bibfnamefont {Muhammad}\
  \bibnamefont {Nadeem}}, \bibinfo {author} {\bibfnamefont {Michael~S}\
  \bibnamefont {Fuhrer}}, \ and\ \bibinfo {author} {\bibfnamefont {Xiaolin}\
  \bibnamefont {Wang}},\ }\bibfield  {title} {\enquote {\bibinfo {title} {The
  superconducting diode effect},}\ }\href
  {https://www.nature.com/articles/s42254-023-00632-w} {\bibfield  {journal}
  {\bibinfo  {journal} {Nature Reviews Physics}\ }\textbf {\bibinfo {volume}
  {5}},\ \bibinfo {pages} {558--577} (\bibinfo {year} {2023})}\BibitemShut
  {NoStop}%
\bibitem [{\citenamefont {Nagaosa}\ and\ \citenamefont
  {Yanase}(2024)}]{NagaosaYanase24}%
  \BibitemOpen
  \bibfield  {author} {\bibinfo {author} {\bibfnamefont {Naoto}\ \bibnamefont
  {Nagaosa}}\ and\ \bibinfo {author} {\bibfnamefont {Youichi}\ \bibnamefont
  {Yanase}},\ }\bibfield  {title} {\enquote {\bibinfo {title} {Nonreciprocal
  {Transport} and {Optical} {Phenomena} in {Quantum} {Materials}},}\ }\href
  {\doibase 10.1146/annurev-conmatphys-032822-033734} {\bibfield  {journal}
  {\bibinfo  {journal} {Annual Review of Condensed Matter Physics}\ }\textbf
  {\bibinfo {volume} {15}},\ \bibinfo {pages} {63--83} (\bibinfo {year}
  {2024})},\ \bibinfo {note} {publisher: Annual Reviews}\BibitemShut {NoStop}%
\bibitem [{\citenamefont {Ma}\ \emph {et~al.}(2025)\citenamefont {Ma},
  \citenamefont {Zhan},\ and\ \citenamefont {Lin}}]{Ma25}%
  \BibitemOpen
  \bibfield  {author} {\bibinfo {author} {\bibfnamefont {Jiajun}\ \bibnamefont
  {Ma}}, \bibinfo {author} {\bibfnamefont {Ruiya}\ \bibnamefont {Zhan}}, \ and\
  \bibinfo {author} {\bibfnamefont {Xiao}\ \bibnamefont {Lin}},\ }\bibfield
  {title} {\enquote {\bibinfo {title} {Superconducting diode effects:
  Mechanisms, materials and applications},}\ }\href {\doibase
  https://doi.org/10.1002/apxr.202400180} {\bibfield  {journal} {\bibinfo
  {journal} {Advanced Physics Research}\ }\textbf {\bibinfo {volume} {n/a}},\
  \bibinfo {pages} {2400180} (\bibinfo {year} {2025})}\BibitemShut {NoStop}%
\bibitem [{\citenamefont {Levitov}\ \emph {et~al.}(1985)\citenamefont
  {Levitov}, \citenamefont {Nazarov},\ and\ \citenamefont
  {Eliashberg}}]{LevitovNazarovEliashberg85}%
  \BibitemOpen
  \bibfield  {author} {\bibinfo {author} {\bibfnamefont {L.~S.}\ \bibnamefont
  {Levitov}}, \bibinfo {author} {\bibfnamefont {Y.~V.}\ \bibnamefont
  {Nazarov}}, \ and\ \bibinfo {author} {\bibfnamefont {G.~M.}\ \bibnamefont
  {Eliashberg}},\ }\bibfield  {title} {\enquote {\bibinfo {title}
  {Magnetostatics of superconductors without an inversion center},}\ }\href
  {https://www.osti.gov/biblio/5259947} {\bibfield  {journal} {\bibinfo
  {journal} {JETP Lett. (Engl. Transl.); (United States)}\ }\textbf {\bibinfo
  {volume} {41:9}} (\bibinfo {year} {1985})},\ \bibinfo {note} {institution: L.
  D. Landau Institute of Theoretical Physics, Academy of Sciences of the
  USSR}\BibitemShut {NoStop}%
\bibitem [{\citenamefont {Edelstein}(1996)}]{Edelstein96}%
  \BibitemOpen
  \bibfield  {author} {\bibinfo {author} {\bibfnamefont {Victor~M}\
  \bibnamefont {Edelstein}},\ }\bibfield  {title} {\enquote {\bibinfo {title}
  {The {G}inzburg - {L}andau equation for superconductors of polar symmetry},}\
  }\href {\doibase 10.1088/0953-8984/8/3/012} {\bibfield  {journal} {\bibinfo
  {journal} {Journal of Physics: Condensed Matter}\ }\textbf {\bibinfo {volume}
  {8}},\ \bibinfo {pages} {339} (\bibinfo {year} {1996})}\BibitemShut {NoStop}%
\bibitem [{\citenamefont {Yuan}\ and\ \citenamefont {Fu}(2022)}]{YuanFu22}%
  \BibitemOpen
  \bibfield  {author} {\bibinfo {author} {\bibfnamefont {Noah F.~Q.}\
  \bibnamefont {Yuan}}\ and\ \bibinfo {author} {\bibfnamefont {Liang}\
  \bibnamefont {Fu}},\ }\bibfield  {title} {\enquote {\bibinfo {title}
  {Supercurrent diode effect and finite-momentum superconductors},}\ }\href
  {\doibase 10.1073/pnas.2119548119} {\bibfield  {journal} {\bibinfo  {journal}
  {Proceedings of the National Academy of Sciences}\ }\textbf {\bibinfo
  {volume} {119}},\ \bibinfo {pages} {e2119548119} (\bibinfo {year} {2022})},\
  \bibinfo {note} {publisher: Proceedings of the National Academy of
  Sciences}\BibitemShut {NoStop}%
\bibitem [{\citenamefont {Ili\ifmmode~\acute{c}\else \'{c}\fi{}}\ and\
  \citenamefont {Bergeret}(2022)}]{IlicBergeret22}%
  \BibitemOpen
  \bibfield  {author} {\bibinfo {author} {\bibfnamefont {S.}~\bibnamefont
  {Ili\ifmmode~\acute{c}\else \'{c}\fi{}}}\ and\ \bibinfo {author}
  {\bibfnamefont {F.~S.}\ \bibnamefont {Bergeret}},\ }\bibfield  {title}
  {\enquote {\bibinfo {title} {Theory of the supercurrent diode effect in
  rashba superconductors with arbitrary disorder},}\ }\href {\doibase
  10.1103/PhysRevLett.128.177001} {\bibfield  {journal} {\bibinfo  {journal}
  {Phys. Rev. Lett.}\ }\textbf {\bibinfo {volume} {128}},\ \bibinfo {pages}
  {177001} (\bibinfo {year} {2022})}\BibitemShut {NoStop}%
\bibitem [{\citenamefont {Davydova}\ \emph {et~al.}(2022)\citenamefont
  {Davydova}, \citenamefont {Prembabu},\ and\ \citenamefont
  {Fu}}]{DavydovaFu22}%
  \BibitemOpen
  \bibfield  {author} {\bibinfo {author} {\bibfnamefont {Margarita}\
  \bibnamefont {Davydova}}, \bibinfo {author} {\bibfnamefont {Saranesh}\
  \bibnamefont {Prembabu}}, \ and\ \bibinfo {author} {\bibfnamefont {Liang}\
  \bibnamefont {Fu}},\ }\bibfield  {title} {\enquote {\bibinfo {title}
  {Universal {Josephson} diode effect},}\ }\href {\doibase
  10.1126/sciadv.abo0309} {\bibfield  {journal} {\bibinfo  {journal} {Science
  Advances}\ }\textbf {\bibinfo {volume} {8}},\ \bibinfo {pages} {eabo0309}
  (\bibinfo {year} {2022})},\ \bibinfo {note} {publisher: American Association
  for the Advancement of Science}\BibitemShut {NoStop}%
\bibitem [{\citenamefont {Hasan}\ \emph {et~al.}(2022)\citenamefont {Hasan},
  \citenamefont {Nesterov}, \citenamefont {Li}, \citenamefont {Houzet},
  \citenamefont {Meyer},\ and\ \citenamefont
  {Levchenko}}]{HasanSongciLevchenko22}%
  \BibitemOpen
  \bibfield  {author} {\bibinfo {author} {\bibfnamefont {Jaglul}\ \bibnamefont
  {Hasan}}, \bibinfo {author} {\bibfnamefont {Konstantin~N.}\ \bibnamefont
  {Nesterov}}, \bibinfo {author} {\bibfnamefont {Songci}\ \bibnamefont {Li}},
  \bibinfo {author} {\bibfnamefont {Manuel}\ \bibnamefont {Houzet}}, \bibinfo
  {author} {\bibfnamefont {Julia~S.}\ \bibnamefont {Meyer}}, \ and\ \bibinfo
  {author} {\bibfnamefont {Alex}\ \bibnamefont {Levchenko}},\ }\bibfield
  {title} {\enquote {\bibinfo {title} {Anomalous {Josephson} effect in planar
  noncentrosymmetric superconducting devices},}\ }\href {\doibase
  10.1103/PhysRevB.106.214518} {\bibfield  {journal} {\bibinfo  {journal}
  {Physical Review B}\ }\textbf {\bibinfo {volume} {106}},\ \bibinfo {pages}
  {214518} (\bibinfo {year} {2022})},\ \bibinfo {note} {publisher: American
  Physical Society}\BibitemShut {NoStop}%
\bibitem [{\citenamefont {Banerjee}\ and\ \citenamefont
  {Scheurer}(2024)}]{BanerjeeScheurer24alt}%
  \BibitemOpen
  \bibfield  {author} {\bibinfo {author} {\bibfnamefont {Sayan}\ \bibnamefont
  {Banerjee}}\ and\ \bibinfo {author} {\bibfnamefont {Mathias~S.}\ \bibnamefont
  {Scheurer}},\ }\bibfield  {title} {\enquote {\bibinfo {title} {Altermagnetic
  superconducting diode effect},}\ }\href {\doibase
  10.1103/PhysRevB.110.024503} {\bibfield  {journal} {\bibinfo  {journal}
  {Phys. Rev. B}\ }\textbf {\bibinfo {volume} {110}},\ \bibinfo {pages}
  {024503} (\bibinfo {year} {2024})}\BibitemShut {NoStop}%
\bibitem [{\citenamefont {Hasan}\ \emph {et~al.}(2024)\citenamefont {Hasan},
  \citenamefont {Shaffer}, \citenamefont {Khodas},\ and\ \citenamefont
  {Levchenko}}]{HasanShafferKhodasLevchenko24}%
  \BibitemOpen
  \bibfield  {author} {\bibinfo {author} {\bibfnamefont {Jaglul}\ \bibnamefont
  {Hasan}}, \bibinfo {author} {\bibfnamefont {Daniel}\ \bibnamefont {Shaffer}},
  \bibinfo {author} {\bibfnamefont {Maxim}\ \bibnamefont {Khodas}}, \ and\
  \bibinfo {author} {\bibfnamefont {Alex}\ \bibnamefont {Levchenko}},\
  }\bibfield  {title} {\enquote {\bibinfo {title} {Supercurrent diode effect in
  helical superconductors},}\ }\href {\doibase 10.1103/PhysRevB.110.024508}
  {\bibfield  {journal} {\bibinfo  {journal} {Physical Review B}\ }\textbf
  {\bibinfo {volume} {110}},\ \bibinfo {pages} {024508} (\bibinfo {year}
  {2024})},\ \bibinfo {note} {publisher: American Physical Society}\BibitemShut
  {NoStop}%
\bibitem [{\citenamefont {Chakraborty}\ and\ \citenamefont
  {Black-Schaffer}(2024)}]{ChakrabortyBlackSchaffer24}%
  \BibitemOpen
  \bibfield  {author} {\bibinfo {author} {\bibfnamefont {Debmalya}\
  \bibnamefont {Chakraborty}}\ and\ \bibinfo {author} {\bibfnamefont
  {Annica~M.}\ \bibnamefont {Black-Schaffer}},\ }\href {\doibase
  10.48550/arXiv.2408.07747} {\enquote {\bibinfo {title} {Perfect
  superconducting diode effect in altermagnets},}\ } (\bibinfo {year} {2024}),\
  \bibinfo {note} {arXiv:2408.07747 [cond-mat]}\BibitemShut {NoStop}%
\bibitem [{\citenamefont {Ili\ifmmode~\acute{c}\else \'{c}\fi{}}\ \emph
  {et~al.}(2024)\citenamefont {Ili\ifmmode~\acute{c}\else \'{c}\fi{}},
  \citenamefont {Virtanen}, \citenamefont {Crawford}, \citenamefont
  {Heikkil\"a},\ and\ \citenamefont {Bergeret}}]{IlicBergeret24}%
  \BibitemOpen
  \bibfield  {author} {\bibinfo {author} {\bibfnamefont {Stefan}\ \bibnamefont
  {Ili\ifmmode~\acute{c}\else \'{c}\fi{}}}, \bibinfo {author} {\bibfnamefont
  {Pauli}\ \bibnamefont {Virtanen}}, \bibinfo {author} {\bibfnamefont {Daniel}\
  \bibnamefont {Crawford}}, \bibinfo {author} {\bibfnamefont {Tero~T.}\
  \bibnamefont {Heikkil\"a}}, \ and\ \bibinfo {author} {\bibfnamefont
  {F.~Sebasti\'an}\ \bibnamefont {Bergeret}},\ }\bibfield  {title} {\enquote
  {\bibinfo {title} {Superconducting diode effect in diffusive superconductors
  and josephson junctions with rashba spin-orbit coupling},}\ }\href {\doibase
  10.1103/PhysRevB.110.L140501} {\bibfield  {journal} {\bibinfo  {journal}
  {Phys. Rev. B}\ }\textbf {\bibinfo {volume} {110}},\ \bibinfo {pages}
  {L140501} (\bibinfo {year} {2024})}\BibitemShut {NoStop}%
\bibitem [{\citenamefont {Ghosh}\ \emph {et~al.}(2024)\citenamefont {Ghosh},
  \citenamefont {Patil}, \citenamefont {Basu}, \citenamefont {Kuldeep},
  \citenamefont {Dutta}, \citenamefont {Jangade}, \citenamefont {Kulkarni},
  \citenamefont {Thamizhavel}, \citenamefont {Steiner}, \citenamefont {von
  Oppen},\ and\ \citenamefont {Deshmukh}}]{Ghosh24}%
  \BibitemOpen
  \bibfield  {author} {\bibinfo {author} {\bibfnamefont {Sanat}\ \bibnamefont
  {Ghosh}}, \bibinfo {author} {\bibfnamefont {Vilas}\ \bibnamefont {Patil}},
  \bibinfo {author} {\bibfnamefont {Amit}\ \bibnamefont {Basu}}, \bibinfo
  {author} {\bibnamefont {Kuldeep}}, \bibinfo {author} {\bibfnamefont
  {Achintya}\ \bibnamefont {Dutta}}, \bibinfo {author} {\bibfnamefont
  {Digambar~A.}\ \bibnamefont {Jangade}}, \bibinfo {author} {\bibfnamefont
  {Ruta}\ \bibnamefont {Kulkarni}}, \bibinfo {author} {\bibfnamefont
  {A.}~\bibnamefont {Thamizhavel}}, \bibinfo {author} {\bibfnamefont
  {Jacob~F.}\ \bibnamefont {Steiner}}, \bibinfo {author} {\bibfnamefont
  {Felix}\ \bibnamefont {von Oppen}}, \ and\ \bibinfo {author} {\bibfnamefont
  {Mandar~M.}\ \bibnamefont {Deshmukh}},\ }\bibfield  {title} {\enquote
  {\bibinfo {title} {High-temperature {Josephson} diode},}\ }\href {\doibase
  10.1038/s41563-024-01804-4} {\bibfield  {journal} {\bibinfo  {journal}
  {Nature Materials}\ ,\ \bibinfo {pages} {1--7}} (\bibinfo {year} {2024})},\
  \bibinfo {note} {publisher: Nature Publishing Group}\BibitemShut {NoStop}%
\bibitem [{\citenamefont {Shaffer}\ \emph {et~al.}(2024)\citenamefont
  {Shaffer}, \citenamefont {Chichinadze},\ and\ \citenamefont
  {Levchenko}}]{ShafferChichinadzeLevchenko24}%
  \BibitemOpen
  \bibfield  {author} {\bibinfo {author} {\bibfnamefont {Daniel}\ \bibnamefont
  {Shaffer}}, \bibinfo {author} {\bibfnamefont {Dmitry~V.}\ \bibnamefont
  {Chichinadze}}, \ and\ \bibinfo {author} {\bibfnamefont {Alex}\ \bibnamefont
  {Levchenko}},\ }\bibfield  {title} {\enquote {\bibinfo {title}
  {Superconducting diode effect in multiphase superconductors},}\ }\href
  {\doibase 10.1103/PhysRevB.110.184509} {\bibfield  {journal} {\bibinfo
  {journal} {Physical Review B}\ }\textbf {\bibinfo {volume} {110}},\ \bibinfo
  {pages} {184509} (\bibinfo {year} {2024})},\ \bibinfo {note} {publisher:
  American Physical Society}\BibitemShut {NoStop}%
\bibitem [{\citenamefont {Hasan}\ \emph {et~al.}(2025)\citenamefont {Hasan},
  \citenamefont {Shaffer}, \citenamefont {Khodas},\ and\ \citenamefont
  {Levchenko}}]{HasanShafferKhodasLevchenko25}%
  \BibitemOpen
  \bibfield  {author} {\bibinfo {author} {\bibfnamefont {Jaglul}\ \bibnamefont
  {Hasan}}, \bibinfo {author} {\bibfnamefont {Daniel}\ \bibnamefont {Shaffer}},
  \bibinfo {author} {\bibfnamefont {Maxim}\ \bibnamefont {Khodas}}, \ and\
  \bibinfo {author} {\bibfnamefont {Alex}\ \bibnamefont {Levchenko}},\ }\href
  {\doibase 10.48550/arXiv.2502.09421} {\enquote {\bibinfo {title}
  {Superconducting diode efficiency from singlet-triplet mixing in disordered
  systems},}\ } (\bibinfo {year} {2025}),\ \bibinfo {note} {arXiv:2502.09421
  [cond-mat]}\BibitemShut {NoStop}%
\bibitem [{\citenamefont {Bankier}\ \emph {et~al.}(2025)\citenamefont
  {Bankier}, \citenamefont {Attias}, \citenamefont {Levchenko},\ and\
  \citenamefont {Khodas}}]{BankierLevchenkoKhodas25}%
  \BibitemOpen
  \bibfield  {author} {\bibinfo {author} {\bibfnamefont {Itai}\ \bibnamefont
  {Bankier}}, \bibinfo {author} {\bibfnamefont {Lotan}\ \bibnamefont {Attias}},
  \bibinfo {author} {\bibfnamefont {Alex}\ \bibnamefont {Levchenko}}, \ and\
  \bibinfo {author} {\bibfnamefont {Maxim}\ \bibnamefont {Khodas}},\ }\href
  {\doibase 10.48550/arXiv.2503.15115} {\enquote {\bibinfo {title}
  {Superconducting diode effect in {Ising} superconductors},}\ } (\bibinfo
  {year} {2025}),\ \bibinfo {note} {arXiv:2503.15115 [cond-mat]}\BibitemShut
  {NoStop}%
\bibitem [{\citenamefont {Krasnov}\ \emph {et~al.}(1997)\citenamefont
  {Krasnov}, \citenamefont {Oboznov},\ and\ \citenamefont
  {Pedersen}}]{Krasnov97}%
  \BibitemOpen
  \bibfield  {author} {\bibinfo {author} {\bibfnamefont {V.~M.}\ \bibnamefont
  {Krasnov}}, \bibinfo {author} {\bibfnamefont {V.~A.}\ \bibnamefont
  {Oboznov}}, \ and\ \bibinfo {author} {\bibfnamefont {N.~F.}\ \bibnamefont
  {Pedersen}},\ }\bibfield  {title} {\enquote {\bibinfo {title} {Fluxon
  dynamics in long {Josephson} junctions in the presence of a temperature
  gradient or spatial nonuniformity},}\ }\href {\doibase
  10.1103/PhysRevB.55.14486} {\bibfield  {journal} {\bibinfo  {journal}
  {Physical Review B}\ }\textbf {\bibinfo {volume} {55}},\ \bibinfo {pages}
  {14486--14498} (\bibinfo {year} {1997})},\ \bibinfo {note} {publisher:
  American Physical Society}\BibitemShut {NoStop}%
\bibitem [{\citenamefont {Zinkl}\ \emph {et~al.}(2022)\citenamefont {Zinkl},
  \citenamefont {Hamamoto},\ and\ \citenamefont {Sigrist}}]{ZinklSigrist22}%
  \BibitemOpen
  \bibfield  {author} {\bibinfo {author} {\bibfnamefont {Bastian}\ \bibnamefont
  {Zinkl}}, \bibinfo {author} {\bibfnamefont {Keita}\ \bibnamefont {Hamamoto}},
  \ and\ \bibinfo {author} {\bibfnamefont {Manfred}\ \bibnamefont {Sigrist}},\
  }\bibfield  {title} {\enquote {\bibinfo {title} {Symmetry conditions for the
  superconducting diode effect in chiral superconductors},}\ }\href {\doibase
  10.1103/PhysRevResearch.4.033167} {\bibfield  {journal} {\bibinfo  {journal}
  {Physical Review Research}\ }\textbf {\bibinfo {volume} {4}},\ \bibinfo
  {pages} {033167} (\bibinfo {year} {2022})},\ \bibinfo {note} {publisher:
  American Physical Society}\BibitemShut {NoStop}%
\bibitem [{\citenamefont {Goldman}\ and\ \citenamefont
  {Kreisman}(1967)}]{Goldman67}%
  \BibitemOpen
  \bibfield  {author} {\bibinfo {author} {\bibfnamefont {A.~M.}\ \bibnamefont
  {Goldman}}\ and\ \bibinfo {author} {\bibfnamefont {P.~J.}\ \bibnamefont
  {Kreisman}},\ }\bibfield  {title} {\enquote {\bibinfo {title} {Meissner
  {Effect} and {Vortex} {Penetration} in {Josephson} {Junctions}},}\ }\href
  {\doibase 10.1103/PhysRev.164.544} {\bibfield  {journal} {\bibinfo  {journal}
  {Physical Review}\ }\textbf {\bibinfo {volume} {164}},\ \bibinfo {pages}
  {544--547} (\bibinfo {year} {1967})},\ \bibinfo {note} {publisher: American
  Physical Society}\BibitemShut {NoStop}%
\bibitem [{\citenamefont {Lee}\ \emph {et~al.}(1999)\citenamefont {Lee},
  \citenamefont {Jank{\'o}}, \citenamefont {Der{\'e}nyi},\ and\ \citenamefont
  {Barab{\'a}si}}]{LeeBarabasi99}%
  \BibitemOpen
  \bibfield  {author} {\bibinfo {author} {\bibfnamefont {C.~S.}\ \bibnamefont
  {Lee}}, \bibinfo {author} {\bibfnamefont {B.}~\bibnamefont {Jank{\'o}}},
  \bibinfo {author} {\bibfnamefont {I.}~\bibnamefont {Der{\'e}nyi}}, \ and\
  \bibinfo {author} {\bibfnamefont {A.~L.}\ \bibnamefont {Barab{\'a}si}},\
  }\bibfield  {title} {\enquote {\bibinfo {title} {Reducing vortex density in
  superconductors using the `ratchet effect'},}\ }\href {\doibase
  10.1038/22485} {\bibfield  {journal} {\bibinfo  {journal} {Nature}\ }\textbf
  {\bibinfo {volume} {400}},\ \bibinfo {pages} {337--340} (\bibinfo {year}
  {1999})}\BibitemShut {NoStop}%
\bibitem [{\citenamefont {Gutfreund}\ \emph {et~al.}(2023)\citenamefont
  {Gutfreund}, \citenamefont {Matsuki}, \citenamefont {Plastovets},
  \citenamefont {Noah}, \citenamefont {Gorzawski}, \citenamefont {Fridman},
  \citenamefont {Yang}, \citenamefont {Buzdin}, \citenamefont {Millo},
  \citenamefont {Robinson},\ and\ \citenamefont {Anahory}}]{GutfreundBuzdin23}%
  \BibitemOpen
  \bibfield  {author} {\bibinfo {author} {\bibfnamefont {Alon}\ \bibnamefont
  {Gutfreund}}, \bibinfo {author} {\bibfnamefont {Hisakazu}\ \bibnamefont
  {Matsuki}}, \bibinfo {author} {\bibfnamefont {Vadim}\ \bibnamefont
  {Plastovets}}, \bibinfo {author} {\bibfnamefont {Avia}\ \bibnamefont {Noah}},
  \bibinfo {author} {\bibfnamefont {Laura}\ \bibnamefont {Gorzawski}}, \bibinfo
  {author} {\bibfnamefont {Nofar}\ \bibnamefont {Fridman}}, \bibinfo {author}
  {\bibfnamefont {Guang}\ \bibnamefont {Yang}}, \bibinfo {author}
  {\bibfnamefont {Alexander}\ \bibnamefont {Buzdin}}, \bibinfo {author}
  {\bibfnamefont {Oded}\ \bibnamefont {Millo}}, \bibinfo {author}
  {\bibfnamefont {Jason W.~A.}\ \bibnamefont {Robinson}}, \ and\ \bibinfo
  {author} {\bibfnamefont {Yonathan}\ \bibnamefont {Anahory}},\ }\bibfield
  {title} {\enquote {\bibinfo {title} {Direct observation of a superconducting
  vortex diode},}\ }\href {\doibase 10.1038/s41467-023-37294-2} {\bibfield
  {journal} {\bibinfo  {journal} {Nature Communications}\ }\textbf {\bibinfo
  {volume} {14}},\ \bibinfo {pages} {1630} (\bibinfo {year} {2023})},\ \bibinfo
  {note} {number: 1 Publisher: Nature Publishing Group}\BibitemShut {NoStop}%
\bibitem [{\citenamefont {Moll}\ and\ \citenamefont
  {Geshkenbein}(2023)}]{MollGeshkenbein23}%
  \BibitemOpen
  \bibfield  {author} {\bibinfo {author} {\bibfnamefont {P.~J.~W.}\
  \bibnamefont {Moll}}\ and\ \bibinfo {author} {\bibfnamefont {V.~B.}\
  \bibnamefont {Geshkenbein}},\ }\bibfield  {title} {\enquote {\bibinfo {title}
  {Evolution of superconducting diodes},}\ }\href {\doibase
  10.1038/s41567-023-02229-7} {\bibfield  {journal} {\bibinfo  {journal}
  {Nature Physics}\ ,\ \bibinfo {pages} {1--2}} (\bibinfo {year} {2023})},\
  \bibinfo {note} {publisher: Nature Publishing Group}\BibitemShut {NoStop}%
\bibitem [{\citenamefont {Gupta}\ \emph {et~al.}(2023)\citenamefont {Gupta},
  \citenamefont {Graziano}, \citenamefont {Pendharkar}, \citenamefont {Dong},
  \citenamefont {Dempsey}, \citenamefont {Palmstr{\o}m},\ and\ \citenamefont
  {Pribiag}}]{GuptaPribiag23}%
  \BibitemOpen
  \bibfield  {author} {\bibinfo {author} {\bibfnamefont {Mohit}\ \bibnamefont
  {Gupta}}, \bibinfo {author} {\bibfnamefont {Gino~V.}\ \bibnamefont
  {Graziano}}, \bibinfo {author} {\bibfnamefont {Mihir}\ \bibnamefont
  {Pendharkar}}, \bibinfo {author} {\bibfnamefont {Jason~T.}\ \bibnamefont
  {Dong}}, \bibinfo {author} {\bibfnamefont {Connor~P.}\ \bibnamefont
  {Dempsey}}, \bibinfo {author} {\bibfnamefont {Chris}\ \bibnamefont
  {Palmstr{\o}m}}, \ and\ \bibinfo {author} {\bibfnamefont {Vlad~S.}\
  \bibnamefont {Pribiag}},\ }\bibfield  {title} {\enquote {\bibinfo {title}
  {Gate-tunable superconducting diode effect in a three-terminal josephson
  device},}\ }\href {\doibase 10.1038/s41467-023-38856-0} {\bibfield  {journal}
  {\bibinfo  {journal} {Nature Communications}\ }\textbf {\bibinfo {volume}
  {14}},\ \bibinfo {pages} {3078} (\bibinfo {year} {2023})}\BibitemShut
  {NoStop}%
\bibitem [{\citenamefont {Chiles}\ \emph {et~al.}(2023)\citenamefont {Chiles},
  \citenamefont {Arnault}, \citenamefont {Chen}, \citenamefont {Larson},
  \citenamefont {Zhao}, \citenamefont {Watanabe}, \citenamefont {Taniguchi},
  \citenamefont {Amet},\ and\ \citenamefont
  {Finkelstein}}]{ChilesFinkelstein23}%
  \BibitemOpen
  \bibfield  {author} {\bibinfo {author} {\bibfnamefont {John}\ \bibnamefont
  {Chiles}}, \bibinfo {author} {\bibfnamefont {Ethan~G.}\ \bibnamefont
  {Arnault}}, \bibinfo {author} {\bibfnamefont {Chun-Chia}\ \bibnamefont
  {Chen}}, \bibinfo {author} {\bibfnamefont {Trevyn F.~Q.}\ \bibnamefont
  {Larson}}, \bibinfo {author} {\bibfnamefont {Lingfei}\ \bibnamefont {Zhao}},
  \bibinfo {author} {\bibfnamefont {Kenji}\ \bibnamefont {Watanabe}}, \bibinfo
  {author} {\bibfnamefont {Takashi}\ \bibnamefont {Taniguchi}}, \bibinfo
  {author} {\bibfnamefont {Fransois}\ \bibnamefont {Amet}}, \ and\ \bibinfo
  {author} {\bibfnamefont {Gleb}\ \bibnamefont {Finkelstein}},\ }\bibfield
  {title} {\enquote {\bibinfo {title} {Nonreciprocal {Supercurrents} in a
  {Field}-{Free} {Graphene} {Josephson} {Triode}},}\ }\href {\doibase
  10.1021/acs.nanolett.3c01276} {\bibfield  {journal} {\bibinfo  {journal}
  {Nano Letters}\ }\textbf {\bibinfo {volume} {23}},\ \bibinfo {pages}
  {5257--5263} (\bibinfo {year} {2023})},\ \bibinfo {note} {publisher: American
  Chemical Society}\BibitemShut {NoStop}%
\bibitem [{\citenamefont {Zhang}\ \emph {et~al.}(2024)\citenamefont {Zhang},
  \citenamefont {Rashid}, \citenamefont {Tanhayi~Ahari}, \citenamefont
  {de~Coster}, \citenamefont {Taniguchi}, \citenamefont {Watanabe},
  \citenamefont {Gilbert}, \citenamefont {Samarth},\ and\ \citenamefont
  {Kayyalha}}]{Zhang24}%
  \BibitemOpen
  \bibfield  {author} {\bibinfo {author} {\bibfnamefont {Fan}\ \bibnamefont
  {Zhang}}, \bibinfo {author} {\bibfnamefont {Asmaul~Smitha}\ \bibnamefont
  {Rashid}}, \bibinfo {author} {\bibfnamefont {Mostafa}\ \bibnamefont
  {Tanhayi~Ahari}}, \bibinfo {author} {\bibfnamefont {George~J.}\ \bibnamefont
  {de~Coster}}, \bibinfo {author} {\bibfnamefont {Takashi}\ \bibnamefont
  {Taniguchi}}, \bibinfo {author} {\bibfnamefont {Kenji}\ \bibnamefont
  {Watanabe}}, \bibinfo {author} {\bibfnamefont {Matthew~J.}\ \bibnamefont
  {Gilbert}}, \bibinfo {author} {\bibfnamefont {Nitin}\ \bibnamefont
  {Samarth}}, \ and\ \bibinfo {author} {\bibfnamefont {Morteza}\ \bibnamefont
  {Kayyalha}},\ }\bibfield  {title} {\enquote {\bibinfo {title}
  {Magnetic-field-free nonreciprocal transport in graphene multiterminal
  josephson junctions},}\ }\href {\doibase 10.1103/PhysRevApplied.21.034011}
  {\bibfield  {journal} {\bibinfo  {journal} {Phys. Rev. Appl.}\ }\textbf
  {\bibinfo {volume} {21}},\ \bibinfo {pages} {034011} (\bibinfo {year}
  {2024})}\BibitemShut {NoStop}%
\bibitem [{\citenamefont {De~Waele}\ \emph {et~al.}(1967)\citenamefont
  {De~Waele}, \citenamefont {Kraan}, \citenamefont {De~Bruyn~Ouboter},\ and\
  \citenamefont {Taconis}}]{DeWaele67}%
  \BibitemOpen
  \bibfield  {author} {\bibinfo {author} {\bibfnamefont {A.~Th. A.~M.}\
  \bibnamefont {De~Waele}}, \bibinfo {author} {\bibfnamefont {W.~H.}\
  \bibnamefont {Kraan}}, \bibinfo {author} {\bibfnamefont {R.}~\bibnamefont
  {De~Bruyn~Ouboter}}, \ and\ \bibinfo {author} {\bibfnamefont {K.~W.}\
  \bibnamefont {Taconis}},\ }\bibfield  {title} {\enquote {\bibinfo {title} {On
  the d.c. voltage across a double point contact between two superconductors at
  zero applied d.c. current in situations in which the junction is in the
  resistive region due to the circulating current of flux quantization},}\
  }\href {\doibase 10.1016/0031-8914(67)90110-3} {\bibfield  {journal}
  {\bibinfo  {journal} {Physica}\ }\textbf {\bibinfo {volume} {37}},\ \bibinfo
  {pages} {114--124} (\bibinfo {year} {1967})}\BibitemShut {NoStop}%
\bibitem [{\citenamefont {Raissi}\ and\ \citenamefont
  {Nordman}(1994)}]{RaissiNordman94}%
  \BibitemOpen
  \bibfield  {author} {\bibinfo {author} {\bibfnamefont {F.}~\bibnamefont
  {Raissi}}\ and\ \bibinfo {author} {\bibfnamefont {J.~E.}\ \bibnamefont
  {Nordman}},\ }\bibfield  {title} {\enquote {\bibinfo {title} {Josephson
  fluxonic diode},}\ }\href {\doibase 10.1063/1.112859} {\bibfield  {journal}
  {\bibinfo  {journal} {Applied Physics Letters}\ }\textbf {\bibinfo {volume}
  {65}},\ \bibinfo {pages} {1838--1840} (\bibinfo {year} {1994})}\BibitemShut
  {NoStop}%
\bibitem [{\citenamefont {Ciaccia}\ \emph {et~al.}(2023)\citenamefont
  {Ciaccia}, \citenamefont {Haller}, \citenamefont {Drachmann}, \citenamefont
  {Lindemann}, \citenamefont {Manfra}, \citenamefont {Schrade},\ and\
  \citenamefont {Sch\"onenberger}}]{CiacciaManfra23}%
  \BibitemOpen
  \bibfield  {author} {\bibinfo {author} {\bibfnamefont {Carlo}\ \bibnamefont
  {Ciaccia}}, \bibinfo {author} {\bibfnamefont {Roy}\ \bibnamefont {Haller}},
  \bibinfo {author} {\bibfnamefont {Asbj\o{}rn C.~C.}\ \bibnamefont
  {Drachmann}}, \bibinfo {author} {\bibfnamefont {Tyler}\ \bibnamefont
  {Lindemann}}, \bibinfo {author} {\bibfnamefont {Michael~J.}\ \bibnamefont
  {Manfra}}, \bibinfo {author} {\bibfnamefont {Constantin}\ \bibnamefont
  {Schrade}}, \ and\ \bibinfo {author} {\bibfnamefont {Christian}\ \bibnamefont
  {Sch\"onenberger}},\ }\bibfield  {title} {\enquote {\bibinfo {title}
  {Gate-tunable josephson diode in proximitized inas supercurrent
  interferometers},}\ }\href {\doibase 10.1103/PhysRevResearch.5.033131}
  {\bibfield  {journal} {\bibinfo  {journal} {Phys. Rev. Res.}\ }\textbf
  {\bibinfo {volume} {5}},\ \bibinfo {pages} {033131} (\bibinfo {year}
  {2023})}\BibitemShut {NoStop}%
\bibitem [{\citenamefont {Reinhardt}\ \emph {et~al.}(2024)\citenamefont
  {Reinhardt}, \citenamefont {Ascherl}, \citenamefont {Costa}, \citenamefont
  {Berger}, \citenamefont {Gronin}, \citenamefont {Gardner}, \citenamefont
  {Lindemann}, \citenamefont {Manfra}, \citenamefont {Fabian}, \citenamefont
  {Kochan}, \citenamefont {Strunk},\ and\ \citenamefont
  {Paradiso}}]{ReinhardtManfra24}%
  \BibitemOpen
  \bibfield  {author} {\bibinfo {author} {\bibfnamefont {S.}~\bibnamefont
  {Reinhardt}}, \bibinfo {author} {\bibfnamefont {T.}~\bibnamefont {Ascherl}},
  \bibinfo {author} {\bibfnamefont {A.}~\bibnamefont {Costa}}, \bibinfo
  {author} {\bibfnamefont {J.}~\bibnamefont {Berger}}, \bibinfo {author}
  {\bibfnamefont {S.}~\bibnamefont {Gronin}}, \bibinfo {author} {\bibfnamefont
  {G.~C.}\ \bibnamefont {Gardner}}, \bibinfo {author} {\bibfnamefont
  {T.}~\bibnamefont {Lindemann}}, \bibinfo {author} {\bibfnamefont {M.~J.}\
  \bibnamefont {Manfra}}, \bibinfo {author} {\bibfnamefont {J.}~\bibnamefont
  {Fabian}}, \bibinfo {author} {\bibfnamefont {D.}~\bibnamefont {Kochan}},
  \bibinfo {author} {\bibfnamefont {C.}~\bibnamefont {Strunk}}, \ and\ \bibinfo
  {author} {\bibfnamefont {N.}~\bibnamefont {Paradiso}},\ }\bibfield  {title}
  {\enquote {\bibinfo {title} {Link between supercurrent diode and anomalous
  josephson effect revealed by gate-controlled interferometry},}\ }\href
  {\doibase 10.1038/s41467-024-48741-z} {\bibfield  {journal} {\bibinfo
  {journal} {Nature Communications}\ }\textbf {\bibinfo {volume} {15}},\
  \bibinfo {pages} {4413} (\bibinfo {year} {2024})}\BibitemShut {NoStop}%
\bibitem [{\citenamefont {Cuozzo}\ \emph {et~al.}(2024)\citenamefont {Cuozzo},
  \citenamefont {Pan}, \citenamefont {Shabani},\ and\ \citenamefont
  {Rossi}}]{CuozzoShabani24}%
  \BibitemOpen
  \bibfield  {author} {\bibinfo {author} {\bibfnamefont {Joseph~J.}\
  \bibnamefont {Cuozzo}}, \bibinfo {author} {\bibfnamefont {Wei}\ \bibnamefont
  {Pan}}, \bibinfo {author} {\bibfnamefont {Javad}\ \bibnamefont {Shabani}}, \
  and\ \bibinfo {author} {\bibfnamefont {Enrico}\ \bibnamefont {Rossi}},\
  }\bibfield  {title} {\enquote {\bibinfo {title} {Microwave-tunable diode
  effect in asymmetric squids with topological josephson junctions},}\ }\href
  {\doibase 10.1103/PhysRevResearch.6.023011} {\bibfield  {journal} {\bibinfo
  {journal} {Phys. Rev. Res.}\ }\textbf {\bibinfo {volume} {6}},\ \bibinfo
  {pages} {023011} (\bibinfo {year} {2024})}\BibitemShut {NoStop}%
\bibitem [{\citenamefont {Davydova}\ \emph {et~al.}(2024)\citenamefont
  {Davydova}, \citenamefont {Geier},\ and\ \citenamefont {Fu}}]{DavydovaFu24}%
  \BibitemOpen
  \bibfield  {author} {\bibinfo {author} {\bibfnamefont {Margarita}\
  \bibnamefont {Davydova}}, \bibinfo {author} {\bibfnamefont {Max}\
  \bibnamefont {Geier}}, \ and\ \bibinfo {author} {\bibfnamefont {Liang}\
  \bibnamefont {Fu}},\ }\bibfield  {title} {\enquote {\bibinfo {title}
  {Nonreciprocal superconductivity},}\ }\href {\doibase 10.1126/sciadv.adr4817}
  {\bibfield  {journal} {\bibinfo  {journal} {Science Advances}\ }\textbf
  {\bibinfo {volume} {10}},\ \bibinfo {pages} {eadr4817} (\bibinfo {year}
  {2024})},\ \bibinfo {note} {publisher: American Association for the
  Advancement of Science}\BibitemShut {NoStop}%
\bibitem [{\citenamefont {Guarcello}\ \emph {et~al.}(2024)\citenamefont
  {Guarcello}, \citenamefont {Pagano},\ and\ \citenamefont
  {Filatrella}}]{GuarcelloFilatrella24}%
  \BibitemOpen
  \bibfield  {author} {\bibinfo {author} {\bibfnamefont {C.}~\bibnamefont
  {Guarcello}}, \bibinfo {author} {\bibfnamefont {S.}~\bibnamefont {Pagano}}, \
  and\ \bibinfo {author} {\bibfnamefont {G.}~\bibnamefont {Filatrella}},\
  }\bibfield  {title} {\enquote {\bibinfo {title} {Efficiency of diode effect
  in asymmetric inline long {Josephson} junctions},}\ }\href {\doibase
  10.1063/5.0211230} {\bibfield  {journal} {\bibinfo  {journal} {Applied
  Physics Letters}\ }\textbf {\bibinfo {volume} {124}},\ \bibinfo {pages}
  {162601} (\bibinfo {year} {2024})}\BibitemShut {NoStop}%
\bibitem [{\citenamefont {Margineda}\ \emph {et~al.}(2025)\citenamefont
  {Margineda}, \citenamefont {Crippa}, \citenamefont {Strambini}, \citenamefont
  {Borgongino}, \citenamefont {Paghi}, \citenamefont {de~Simoni}, \citenamefont
  {Sorba}, \citenamefont {Fukaya}, \citenamefont {Mercaldo}, \citenamefont
  {Ortix}, \citenamefont {Cuoco},\ and\ \citenamefont
  {Giazotto}}]{MarginedaGiazotto25}%
  \BibitemOpen
  \bibfield  {author} {\bibinfo {author} {\bibfnamefont {Daniel}\ \bibnamefont
  {Margineda}}, \bibinfo {author} {\bibfnamefont {Alessandro}\ \bibnamefont
  {Crippa}}, \bibinfo {author} {\bibfnamefont {Elia}\ \bibnamefont
  {Strambini}}, \bibinfo {author} {\bibfnamefont {Laura}\ \bibnamefont
  {Borgongino}}, \bibinfo {author} {\bibfnamefont {Alessandro}\ \bibnamefont
  {Paghi}}, \bibinfo {author} {\bibfnamefont {Giorgio}\ \bibnamefont
  {de~Simoni}}, \bibinfo {author} {\bibfnamefont {Lucia}\ \bibnamefont
  {Sorba}}, \bibinfo {author} {\bibfnamefont {Yuri}\ \bibnamefont {Fukaya}},
  \bibinfo {author} {\bibfnamefont {Maria~Teresa}\ \bibnamefont {Mercaldo}},
  \bibinfo {author} {\bibfnamefont {Carmine}\ \bibnamefont {Ortix}}, \bibinfo
  {author} {\bibfnamefont {Mario}\ \bibnamefont {Cuoco}}, \ and\ \bibinfo
  {author} {\bibfnamefont {Francesco}\ \bibnamefont {Giazotto}},\ }\bibfield
  {title} {\enquote {\bibinfo {title} {Back-action supercurrent rectifiers},}\
  }\href {\doibase 10.1038/s42005-024-01931-z} {\bibfield  {journal} {\bibinfo
  {journal} {Communications Physics}\ }\textbf {\bibinfo {volume} {8}},\
  \bibinfo {pages} {1--7} (\bibinfo {year} {2025})},\ \bibinfo {note}
  {publisher: Nature Publishing Group}\BibitemShut {NoStop}%
\bibitem [{\citenamefont {Cuozzo}\ and\ \citenamefont
  {L{\'e}onard}(2025)}]{Cuozzo25}%
  \BibitemOpen
  \bibfield  {author} {\bibinfo {author} {\bibfnamefont {Joseph~J.}\
  \bibnamefont {Cuozzo}}\ and\ \bibinfo {author} {\bibfnamefont {Fran{\c
  c}ois}\ \bibnamefont {L{\'e}onard}},\ }\href
  {https://arxiv.org/abs/2504.02948} {\enquote {\bibinfo {title} {Perfect
  supercurrent diode efficiency in chiral nanotube-based josephson
  junctions},}\ } (\bibinfo {year} {2025}),\ \Eprint
  {http://arxiv.org/abs/2504.02948} {arXiv:2504.02948 [cond-mat.supr-con]}
  \BibitemShut {NoStop}%
\bibitem [{\citenamefont {Edwards}\ and\ \citenamefont
  {Newhouse}(1962)}]{EdwardsNewhouse62}%
  \BibitemOpen
  \bibfield  {author} {\bibinfo {author} {\bibfnamefont {H.~H.}\ \bibnamefont
  {Edwards}}\ and\ \bibinfo {author} {\bibfnamefont {V.~L.}\ \bibnamefont
  {Newhouse}},\ }\bibfield  {title} {\enquote {\bibinfo {title}
  {Superconducting {Film} {Geometry} {With} {Strong} {Critical} {Current}
  {Asymmetry}},}\ }\href {\doibase 10.1063/1.1777183} {\bibfield  {journal}
  {\bibinfo  {journal} {Journal of Applied Physics}\ }\textbf {\bibinfo
  {volume} {33}},\ \bibinfo {pages} {868--874} (\bibinfo {year}
  {1962})}\BibitemShut {NoStop}%
\bibitem [{\citenamefont {Banerjee}\ and\ \citenamefont
  {Scheurer}(2025)}]{BanerjeeScheurer25}%
  \BibitemOpen
  \bibfield  {author} {\bibinfo {author} {\bibfnamefont {Sayan}\ \bibnamefont
  {Banerjee}}\ and\ \bibinfo {author} {\bibfnamefont {Mathias~S.}\ \bibnamefont
  {Scheurer}},\ }\href {\doibase 10.48550/arXiv.2501.01501} {\enquote {\bibinfo
  {title} {Dissipation-enhanced non-reciprocal superconductivity: application
  to multi-valley superconductors},}\ } (\bibinfo {year} {2025}),\ \bibinfo
  {note} {arXiv:2501.01501 [cond-mat]}\BibitemShut {NoStop}%
\bibitem [{\citenamefont {Daido}\ and\ \citenamefont
  {Yanase}(2025)}]{DaidoYanase25}%
  \BibitemOpen
  \bibfield  {author} {\bibinfo {author} {\bibfnamefont {Akito}\ \bibnamefont
  {Daido}}\ and\ \bibinfo {author} {\bibfnamefont {Youichi}\ \bibnamefont
  {Yanase}},\ }\bibfield  {title} {\enquote {\bibinfo {title} {Unidirectional
  superconductivity and superconducting diode effect induced by dissipation},}\
  }\href {\doibase 10.1103/PhysRevB.111.L020508} {\bibfield  {journal}
  {\bibinfo  {journal} {Physical Review B}\ }\textbf {\bibinfo {volume}
  {111}},\ \bibinfo {pages} {L020508} (\bibinfo {year} {2025})},\ \bibinfo
  {note} {publisher: American Physical Society}\BibitemShut {NoStop}%
\bibitem [{\citenamefont {Menditto}\ \emph {et~al.}(2016)\citenamefont
  {Menditto}, \citenamefont {Sickinger}, \citenamefont {Weides}, \citenamefont
  {Kohlstedt}, \citenamefont {Koelle}, \citenamefont {Kleiner},\ and\
  \citenamefont {Goldobin}}]{MendittoGoldobin16}%
  \BibitemOpen
  \bibfield  {author} {\bibinfo {author} {\bibfnamefont {R.}~\bibnamefont
  {Menditto}}, \bibinfo {author} {\bibfnamefont {H.}~\bibnamefont {Sickinger}},
  \bibinfo {author} {\bibfnamefont {M.}~\bibnamefont {Weides}}, \bibinfo
  {author} {\bibfnamefont {H.}~\bibnamefont {Kohlstedt}}, \bibinfo {author}
  {\bibfnamefont {D.}~\bibnamefont {Koelle}}, \bibinfo {author} {\bibfnamefont
  {R.}~\bibnamefont {Kleiner}}, \ and\ \bibinfo {author} {\bibfnamefont
  {E.}~\bibnamefont {Goldobin}},\ }\bibfield  {title} {\enquote {\bibinfo
  {title} {Tunable $\varphi$ {Josephson} junction ratchet},}\ }\href {\doibase
  10.1103/PhysRevE.94.042202} {\bibfield  {journal} {\bibinfo  {journal}
  {Physical Review E}\ }\textbf {\bibinfo {volume} {94}},\ \bibinfo {pages}
  {042202} (\bibinfo {year} {2016})},\ \bibinfo {note} {publisher: American
  Physical Society}\BibitemShut {NoStop}%
\bibitem [{\citenamefont {Schmid}\ \emph {et~al.}(2024)\citenamefont {Schmid},
  \citenamefont {Jozani}, \citenamefont {Kleiner}, \citenamefont {Koelle},\
  and\ \citenamefont {Goldobin}}]{SchmidGoldobin24}%
  \BibitemOpen
  \bibfield  {author} {\bibinfo {author} {\bibfnamefont {Christoph}\
  \bibnamefont {Schmid}}, \bibinfo {author} {\bibfnamefont {Alireza}\
  \bibnamefont {Jozani}}, \bibinfo {author} {\bibfnamefont {Reinhold}\
  \bibnamefont {Kleiner}}, \bibinfo {author} {\bibfnamefont {Dieter}\
  \bibnamefont {Koelle}}, \ and\ \bibinfo {author} {\bibfnamefont {Edward}\
  \bibnamefont {Goldobin}},\ }\href {\doibase 10.48550/arXiv.2408.01521}
  {\enquote {\bibinfo {title} {{YB}a$_2${C}u$_3${O}$_7$ josephson diode
  operating as a high-efficiency ratchet},}\ } (\bibinfo {year} {2024}),\
  \bibinfo {note} {arXiv:2408.01521 [cond-mat, physics:physics]}\BibitemShut
  {NoStop}%
\bibitem [{\citenamefont {Su}\ \emph {et~al.}(2024)\citenamefont {Su},
  \citenamefont {Wang}, \citenamefont {Gao}, \citenamefont {Luo}, \citenamefont
  {Yan}, \citenamefont {Wu}, \citenamefont {Li}, \citenamefont {Shen},
  \citenamefont {Lu}, \citenamefont {Pan}, \citenamefont {Zhao}, \citenamefont
  {Zhang},\ and\ \citenamefont {Xu}}]{Su24}%
  \BibitemOpen
  \bibfield  {author} {\bibinfo {author} {\bibfnamefont {Haitian}\ \bibnamefont
  {Su}}, \bibinfo {author} {\bibfnamefont {Ji-Yin}\ \bibnamefont {Wang}},
  \bibinfo {author} {\bibfnamefont {Han}\ \bibnamefont {Gao}}, \bibinfo
  {author} {\bibfnamefont {Yi}~\bibnamefont {Luo}}, \bibinfo {author}
  {\bibfnamefont {Shili}\ \bibnamefont {Yan}}, \bibinfo {author} {\bibfnamefont
  {Xingjun}\ \bibnamefont {Wu}}, \bibinfo {author} {\bibfnamefont {Guoan}\
  \bibnamefont {Li}}, \bibinfo {author} {\bibfnamefont {Jie}\ \bibnamefont
  {Shen}}, \bibinfo {author} {\bibfnamefont {Li}~\bibnamefont {Lu}}, \bibinfo
  {author} {\bibfnamefont {Dong}\ \bibnamefont {Pan}}, \bibinfo {author}
  {\bibfnamefont {Jianhua}\ \bibnamefont {Zhao}}, \bibinfo {author}
  {\bibfnamefont {Po}~\bibnamefont {Zhang}}, \ and\ \bibinfo {author}
  {\bibfnamefont {H.~Q.}\ \bibnamefont {Xu}},\ }\bibfield  {title} {\enquote
  {\bibinfo {title} {Microwave-assisted unidirectional superconductivity in
  al-inas nanowire-al junctions under magnetic fields},}\ }\href {\doibase
  10.1103/PhysRevLett.133.087001} {\bibfield  {journal} {\bibinfo  {journal}
  {Phys. Rev. Lett.}\ }\textbf {\bibinfo {volume} {133}},\ \bibinfo {pages}
  {087001} (\bibinfo {year} {2024})}\BibitemShut {NoStop}%
\bibitem [{\citenamefont {Matsuo}\ \emph {et~al.}(2025)\citenamefont {Matsuo},
  \citenamefont {Deacon}, \citenamefont {Kobayashi}, \citenamefont {Sato},
  \citenamefont {Yokoyama}, \citenamefont {Lindemann}, \citenamefont {Gronin},
  \citenamefont {Gardner}, \citenamefont {Ishibashi}, \citenamefont {Manfra},\
  and\ \citenamefont {Tarucha}}]{MatsuoManfra25}%
  \BibitemOpen
  \bibfield  {author} {\bibinfo {author} {\bibfnamefont {Sadashige}\
  \bibnamefont {Matsuo}}, \bibinfo {author} {\bibfnamefont {Russell~S.}\
  \bibnamefont {Deacon}}, \bibinfo {author} {\bibfnamefont {Shohei}\
  \bibnamefont {Kobayashi}}, \bibinfo {author} {\bibfnamefont {Yosuke}\
  \bibnamefont {Sato}}, \bibinfo {author} {\bibfnamefont {Tomohiro}\
  \bibnamefont {Yokoyama}}, \bibinfo {author} {\bibfnamefont {Tyler}\
  \bibnamefont {Lindemann}}, \bibinfo {author} {\bibfnamefont {Sergei}\
  \bibnamefont {Gronin}}, \bibinfo {author} {\bibfnamefont {Geoffrey~C.}\
  \bibnamefont {Gardner}}, \bibinfo {author} {\bibfnamefont {Koji}\
  \bibnamefont {Ishibashi}}, \bibinfo {author} {\bibfnamefont {Michael~J.}\
  \bibnamefont {Manfra}}, \ and\ \bibinfo {author} {\bibfnamefont {Seigo}\
  \bibnamefont {Tarucha}},\ }\bibfield  {title} {\enquote {\bibinfo {title}
  {Shapiro response of superconducting diode effect derived from {Andreev}
  molecules},}\ }\href {\doibase 10.1103/PhysRevB.111.094512} {\bibfield
  {journal} {\bibinfo  {journal} {Physical Review B}\ }\textbf {\bibinfo
  {volume} {111}},\ \bibinfo {pages} {094512} (\bibinfo {year} {2025})},\
  \bibinfo {note} {publisher: American Physical Society}\BibitemShut {NoStop}%
\bibitem [{\citenamefont {Borgongino}\ \emph {et~al.}(2025)\citenamefont
  {Borgongino}, \citenamefont {Souto}, \citenamefont {Paghi}, \citenamefont
  {Senesi}, \citenamefont {Skibinska}, \citenamefont {Sorba}, \citenamefont
  {Giazotto},\ and\ \citenamefont {Strambini}}]{BorgonginoGiazotto25}%
  \BibitemOpen
  \bibfield  {author} {\bibinfo {author} {\bibfnamefont {L.}~\bibnamefont
  {Borgongino}}, \bibinfo {author} {\bibfnamefont {R.~Seoane}\ \bibnamefont
  {Souto}}, \bibinfo {author} {\bibfnamefont {A.}~\bibnamefont {Paghi}},
  \bibinfo {author} {\bibfnamefont {G.}~\bibnamefont {Senesi}}, \bibinfo
  {author} {\bibfnamefont {K.}~\bibnamefont {Skibinska}}, \bibinfo {author}
  {\bibfnamefont {L.}~\bibnamefont {Sorba}}, \bibinfo {author} {\bibfnamefont
  {F.}~\bibnamefont {Giazotto}}, \ and\ \bibinfo {author} {\bibfnamefont
  {E.}~\bibnamefont {Strambini}},\ }\href {\doibase 10.48550/arXiv.2504.08691}
  {\enquote {\bibinfo {title} {Biharmonic-drive tunable {Josephson} diode},}\ }
  (\bibinfo {year} {2025}),\ \bibinfo {note} {arXiv:2504.08691
  [cond-mat]}\BibitemShut {NoStop}%
\bibitem [{\citenamefont {Seoane~Souto}\ \emph {et~al.}(2024)\citenamefont
  {Seoane~Souto}, \citenamefont {Leijnse}, \citenamefont {Schrade},
  \citenamefont {Valentini}, \citenamefont {Katsaros},\ and\ \citenamefont
  {Danon}}]{SeoaneSoutoDanon24}%
  \BibitemOpen
  \bibfield  {author} {\bibinfo {author} {\bibfnamefont {Rub\'en}\ \bibnamefont
  {Seoane~Souto}}, \bibinfo {author} {\bibfnamefont {Martin}\ \bibnamefont
  {Leijnse}}, \bibinfo {author} {\bibfnamefont {Constantin}\ \bibnamefont
  {Schrade}}, \bibinfo {author} {\bibfnamefont {Marco}\ \bibnamefont
  {Valentini}}, \bibinfo {author} {\bibfnamefont {Georgios}\ \bibnamefont
  {Katsaros}}, \ and\ \bibinfo {author} {\bibfnamefont {Jeroen}\ \bibnamefont
  {Danon}},\ }\bibfield  {title} {\enquote {\bibinfo {title} {Tuning the
  {Josephson} diode response with an ac current},}\ }\href {\doibase
  10.1103/PhysRevResearch.6.L022002} {\bibfield  {journal} {\bibinfo  {journal}
  {Physical Review Research}\ }\textbf {\bibinfo {volume} {6}},\ \bibinfo
  {pages} {L022002} (\bibinfo {year} {2024})},\ \bibinfo {note} {publisher:
  American Physical Society}\BibitemShut {NoStop}%
\bibitem [{\citenamefont {Monroe}\ \emph {et~al.}(2024)\citenamefont {Monroe},
  \citenamefont {Shen}, \citenamefont {Tringali}, \citenamefont {Alidoust},
  \citenamefont {Zhou},\ and\ \citenamefont {{\v{Z}}uti{\'c}}}]{MonroeZutic24}%
  \BibitemOpen
  \bibfield  {author} {\bibinfo {author} {\bibfnamefont {David}\ \bibnamefont
  {Monroe}}, \bibinfo {author} {\bibfnamefont {Chenghao}\ \bibnamefont {Shen}},
  \bibinfo {author} {\bibfnamefont {Dario}\ \bibnamefont {Tringali}}, \bibinfo
  {author} {\bibfnamefont {Mohammad}\ \bibnamefont {Alidoust}}, \bibinfo
  {author} {\bibfnamefont {Tong}\ \bibnamefont {Zhou}}, \ and\ \bibinfo
  {author} {\bibfnamefont {Igor}\ \bibnamefont {{\v{Z}}uti{\'c}}},\ }\bibfield
  {title} {\enquote {\bibinfo {title} {Phase jumps in josephson junctions with
  time-dependent spin--orbit coupling},}\ }\href@noop {} {\bibfield  {journal}
  {\bibinfo  {journal} {Applied Physics Letters}\ }\textbf {\bibinfo {volume}
  {125}} (\bibinfo {year} {2024})}\BibitemShut {NoStop}%
\bibitem [{\citenamefont {Soori}(2023)}]{Soori23I}%
  \BibitemOpen
  \bibfield  {author} {\bibinfo {author} {\bibfnamefont {Abhiram}\ \bibnamefont
  {Soori}},\ }\bibfield  {title} {\enquote {\bibinfo {title} {Nonequilibrium
  {Josephson} diode effect in periodically driven {SNS} junctions},}\ }\href
  {\doibase 10.1088/1402-4896/acd02f} {\bibfield  {journal} {\bibinfo
  {journal} {Physica Scripta}\ }\textbf {\bibinfo {volume} {98}},\ \bibinfo
  {pages} {065917} (\bibinfo {year} {2023})},\ \bibinfo {note} {publisher: IOP
  Publishing}\BibitemShut {NoStop}%
\bibitem [{\citenamefont {Ortega-Taberner}\ \emph {et~al.}(2023)\citenamefont
  {Ortega-Taberner}, \citenamefont {Jauho},\ and\ \citenamefont
  {Paaske}}]{OrtegaTaberner23}%
  \BibitemOpen
  \bibfield  {author} {\bibinfo {author} {\bibfnamefont {Carlos}\ \bibnamefont
  {Ortega-Taberner}}, \bibinfo {author} {\bibfnamefont {Antti-Pekka}\
  \bibnamefont {Jauho}}, \ and\ \bibinfo {author} {\bibfnamefont {Jens}\
  \bibnamefont {Paaske}},\ }\bibfield  {title} {\enquote {\bibinfo {title}
  {Anomalous {Josephson} current through a driven double quantum dot},}\ }\href
  {\doibase 10.1103/PhysRevB.107.115165} {\bibfield  {journal} {\bibinfo
  {journal} {Physical Review B}\ }\textbf {\bibinfo {volume} {107}},\ \bibinfo
  {pages} {115165} (\bibinfo {year} {2023})},\ \bibinfo {note} {publisher:
  American Physical Society}\BibitemShut {NoStop}%
\bibitem [{\citenamefont {Liu}\ \emph {et~al.}(2024)\citenamefont {Liu},
  \citenamefont {Smith}, \citenamefont {Andreev},\ and\ \citenamefont
  {Spivak}}]{LiuAndreevSpivak24}%
  \BibitemOpen
  \bibfield  {author} {\bibinfo {author} {\bibfnamefont {T.}~\bibnamefont
  {Liu}}, \bibinfo {author} {\bibfnamefont {M.}~\bibnamefont {Smith}}, \bibinfo
  {author} {\bibfnamefont {A.~V.}\ \bibnamefont {Andreev}}, \ and\ \bibinfo
  {author} {\bibfnamefont {B.~Z.}\ \bibnamefont {Spivak}},\ }\bibfield  {title}
  {\enquote {\bibinfo {title} {Giant nonreciprocity of current-voltage
  characteristics of noncentrosymmetric superconductor--normal
  metal--superconductor junctions},}\ }\href {\doibase
  10.1103/PhysRevB.109.L020501} {\bibfield  {journal} {\bibinfo  {journal}
  {Physical Review B}\ }\textbf {\bibinfo {volume} {109}},\ \bibinfo {pages}
  {L020501} (\bibinfo {year} {2024})},\ \bibinfo {note} {publisher: American
  Physical Society}\BibitemShut {NoStop}%
\bibitem [{\citenamefont {Zazunov}\ \emph {et~al.}(2024)\citenamefont
  {Zazunov}, \citenamefont {Rech}, \citenamefont {Jonckheere}, \citenamefont
  {Gr\'emaud}, \citenamefont {Martin},\ and\ \citenamefont
  {Egger}}]{ZazunovEgger24}%
  \BibitemOpen
  \bibfield  {author} {\bibinfo {author} {\bibfnamefont {A.}~\bibnamefont
  {Zazunov}}, \bibinfo {author} {\bibfnamefont {J.}~\bibnamefont {Rech}},
  \bibinfo {author} {\bibfnamefont {T.}~\bibnamefont {Jonckheere}}, \bibinfo
  {author} {\bibfnamefont {B.}~\bibnamefont {Gr\'emaud}}, \bibinfo {author}
  {\bibfnamefont {T.}~\bibnamefont {Martin}}, \ and\ \bibinfo {author}
  {\bibfnamefont {R.}~\bibnamefont {Egger}},\ }\bibfield  {title} {\enquote
  {\bibinfo {title} {Nonreciprocal charge transport and subharmonic structure
  in voltage-biased josephson diodes},}\ }\href {\doibase
  10.1103/PhysRevB.109.024504} {\bibfield  {journal} {\bibinfo  {journal}
  {Phys. Rev. B}\ }\textbf {\bibinfo {volume} {109}},\ \bibinfo {pages}
  {024504} (\bibinfo {year} {2024})}\BibitemShut {NoStop}%
\bibitem [{\citenamefont {Zapata}\ \emph {et~al.}(1996)\citenamefont {Zapata},
  \citenamefont {Bartussek}, \citenamefont {Sols},\ and\ \citenamefont
  {H\"anggi}}]{Zapata96}%
  \BibitemOpen
  \bibfield  {author} {\bibinfo {author} {\bibfnamefont {I.}~\bibnamefont
  {Zapata}}, \bibinfo {author} {\bibfnamefont {R.}~\bibnamefont {Bartussek}},
  \bibinfo {author} {\bibfnamefont {F.}~\bibnamefont {Sols}}, \ and\ \bibinfo
  {author} {\bibfnamefont {P.}~\bibnamefont {H\"anggi}},\ }\bibfield  {title}
  {\enquote {\bibinfo {title} {Voltage rectification by a squid ratchet},}\
  }\href {\doibase 10.1103/PhysRevLett.77.2292} {\bibfield  {journal} {\bibinfo
   {journal} {Phys. Rev. Lett.}\ }\textbf {\bibinfo {volume} {77}},\ \bibinfo
  {pages} {2292--2295} (\bibinfo {year} {1996})}\BibitemShut {NoStop}%
\bibitem [{\citenamefont {Goldobin}\ \emph {et~al.}(2001)\citenamefont
  {Goldobin}, \citenamefont {Sterck},\ and\ \citenamefont
  {Koelle}}]{Goldobin01}%
  \BibitemOpen
  \bibfield  {author} {\bibinfo {author} {\bibfnamefont {E.}~\bibnamefont
  {Goldobin}}, \bibinfo {author} {\bibfnamefont {A.}~\bibnamefont {Sterck}}, \
  and\ \bibinfo {author} {\bibfnamefont {D.}~\bibnamefont {Koelle}},\
  }\bibfield  {title} {\enquote {\bibinfo {title} {Josephson vortex in a
  ratchet potential: {Theory}},}\ }\href {\doibase 10.1103/PhysRevE.63.031111}
  {\bibfield  {journal} {\bibinfo  {journal} {Physical Review E}\ }\textbf
  {\bibinfo {volume} {63}},\ \bibinfo {pages} {031111} (\bibinfo {year}
  {2001})},\ \bibinfo {note} {publisher: American Physical Society}\BibitemShut
  {NoStop}%
\bibitem [{\citenamefont {Carapella}\ and\ \citenamefont
  {Costabile}(2001)}]{Carapella01}%
  \BibitemOpen
  \bibfield  {author} {\bibinfo {author} {\bibfnamefont {G.}~\bibnamefont
  {Carapella}}\ and\ \bibinfo {author} {\bibfnamefont {G.}~\bibnamefont
  {Costabile}},\ }\bibfield  {title} {\enquote {\bibinfo {title} {Ratchet
  {Effect}: {Demonstration} of a {Relativistic} {Fluxon} {Diode}},}\ }\href
  {\doibase 10.1103/PhysRevLett.87.077002} {\bibfield  {journal} {\bibinfo
  {journal} {Physical Review Letters}\ }\textbf {\bibinfo {volume} {87}},\
  \bibinfo {pages} {077002} (\bibinfo {year} {2001})},\ \bibinfo {note}
  {publisher: American Physical Society}\BibitemShut {NoStop}%
\bibitem [{\citenamefont {Lee}(2003)}]{Lee03}%
  \BibitemOpen
  \bibfield  {author} {\bibinfo {author} {\bibfnamefont {K.~H.}\ \bibnamefont
  {Lee}},\ }\bibfield  {title} {\enquote {\bibinfo {title} {Ratchet effect in
  an ac-current driven {Josephson} junction array},}\ }\href {\doibase
  10.1063/1.1591244} {\bibfield  {journal} {\bibinfo  {journal} {Applied
  Physics Letters}\ }\textbf {\bibinfo {volume} {83}},\ \bibinfo {pages}
  {117--119} (\bibinfo {year} {2003})}\BibitemShut {NoStop}%
\bibitem [{\citenamefont {Golod}\ and\ \citenamefont
  {Krasnov}(2022)}]{GolodKrasnov22}%
  \BibitemOpen
  \bibfield  {author} {\bibinfo {author} {\bibfnamefont {Taras}\ \bibnamefont
  {Golod}}\ and\ \bibinfo {author} {\bibfnamefont {Vladimir~M.}\ \bibnamefont
  {Krasnov}},\ }\bibfield  {title} {\enquote {\bibinfo {title} {Demonstration
  of a superconducting diode-with-memory, operational at zero magnetic field
  with switchable nonreciprocity},}\ }\href {\doibase
  10.1038/s41467-022-31256-w} {\bibfield  {journal} {\bibinfo  {journal}
  {Nature Communications}\ }\textbf {\bibinfo {volume} {13}},\ \bibinfo {pages}
  {3658} (\bibinfo {year} {2022})},\ \bibinfo {note} {number: 1 Publisher:
  Nature Publishing Group}\BibitemShut {NoStop}%
\bibitem [{\citenamefont {Souto}\ \emph {et~al.}(2022)\citenamefont {Souto},
  \citenamefont {Leijnse},\ and\ \citenamefont {Schrade}}]{SeoaneSouto22}%
  \BibitemOpen
  \bibfield  {author} {\bibinfo {author} {\bibfnamefont {Rub\'en~Seoane}\
  \bibnamefont {Souto}}, \bibinfo {author} {\bibfnamefont {Martin}\
  \bibnamefont {Leijnse}}, \ and\ \bibinfo {author} {\bibfnamefont
  {Constantin}\ \bibnamefont {Schrade}},\ }\bibfield  {title} {\enquote
  {\bibinfo {title} {Josephson diode effect in supercurrent interferometers},}\
  }\href {\doibase 10.1103/PhysRevLett.129.267702} {\bibfield  {journal}
  {\bibinfo  {journal} {Phys. Rev. Lett.}\ }\textbf {\bibinfo {volume} {129}},\
  \bibinfo {pages} {267702} (\bibinfo {year} {2022})}\BibitemShut {NoStop}%
\bibitem [{\citenamefont {Valentini}\ \emph {et~al.}(2024)\citenamefont
  {Valentini}, \citenamefont {Sagi}, \citenamefont {Baghumyan}, \citenamefont
  {de~Gijsel}, \citenamefont {Jung}, \citenamefont {Calcaterra}, \citenamefont
  {Ballabio}, \citenamefont {Aguilera~Servin}, \citenamefont {Aggarwal},
  \citenamefont {Janik}, \citenamefont {Adletzberger}, \citenamefont
  {Seoane~Souto}, \citenamefont {Leijnse}, \citenamefont {Danon}, \citenamefont
  {Schrade}, \citenamefont {Bakkers}, \citenamefont {Chrastina}, \citenamefont
  {Isella},\ and\ \citenamefont {Katsaros}}]{ValentiniDanon24}%
  \BibitemOpen
  \bibfield  {author} {\bibinfo {author} {\bibfnamefont {Marco}\ \bibnamefont
  {Valentini}}, \bibinfo {author} {\bibfnamefont {Oliver}\ \bibnamefont
  {Sagi}}, \bibinfo {author} {\bibfnamefont {Levon}\ \bibnamefont {Baghumyan}},
  \bibinfo {author} {\bibfnamefont {Thijs}\ \bibnamefont {de~Gijsel}}, \bibinfo
  {author} {\bibfnamefont {Jason}\ \bibnamefont {Jung}}, \bibinfo {author}
  {\bibfnamefont {Stefano}\ \bibnamefont {Calcaterra}}, \bibinfo {author}
  {\bibfnamefont {Andrea}\ \bibnamefont {Ballabio}}, \bibinfo {author}
  {\bibfnamefont {Juan}\ \bibnamefont {Aguilera~Servin}}, \bibinfo {author}
  {\bibfnamefont {Kushagra}\ \bibnamefont {Aggarwal}}, \bibinfo {author}
  {\bibfnamefont {Marian}\ \bibnamefont {Janik}}, \bibinfo {author}
  {\bibfnamefont {Thomas}\ \bibnamefont {Adletzberger}}, \bibinfo {author}
  {\bibfnamefont {Rub\'en}\ \bibnamefont {Seoane~Souto}}, \bibinfo {author}
  {\bibfnamefont {Martin}\ \bibnamefont {Leijnse}}, \bibinfo {author}
  {\bibfnamefont {Jeroen}\ \bibnamefont {Danon}}, \bibinfo {author}
  {\bibfnamefont {Constantin}\ \bibnamefont {Schrade}}, \bibinfo {author}
  {\bibfnamefont {Erik}\ \bibnamefont {Bakkers}}, \bibinfo {author}
  {\bibfnamefont {Daniel}\ \bibnamefont {Chrastina}}, \bibinfo {author}
  {\bibfnamefont {Giovanni}\ \bibnamefont {Isella}}, \ and\ \bibinfo {author}
  {\bibfnamefont {Georgios}\ \bibnamefont {Katsaros}},\ }\bibfield  {title}
  {\enquote {\bibinfo {title} {Parity-conserving {Cooper}-pair transport and
  ideal superconducting diode in planar germanium},}\ }\href {\doibase
  10.1038/s41467-023-44114-0} {\bibfield  {journal} {\bibinfo  {journal}
  {Nature Communications}\ }\textbf {\bibinfo {volume} {15}},\ \bibinfo {pages}
  {169} (\bibinfo {year} {2024})},\ \bibinfo {note} {publisher: Nature
  Publishing Group}\BibitemShut {NoStop}%
\bibitem [{\citenamefont {Trahms}\ \emph {et~al.}(2023)\citenamefont {Trahms},
  \citenamefont {Melischek}, \citenamefont {Steiner}, \citenamefont {Mahendru},
  \citenamefont {Tamir}, \citenamefont {Bogdanoff}, \citenamefont {Peters},
  \citenamefont {Reecht}, \citenamefont {Winkelmann}, \citenamefont {von
  Oppen},\ and\ \citenamefont {Franke}}]{TrahmsVonOppen23}%
  \BibitemOpen
  \bibfield  {author} {\bibinfo {author} {\bibfnamefont {Martina}\ \bibnamefont
  {Trahms}}, \bibinfo {author} {\bibfnamefont {Larissa}\ \bibnamefont
  {Melischek}}, \bibinfo {author} {\bibfnamefont {Jacob~F.}\ \bibnamefont
  {Steiner}}, \bibinfo {author} {\bibfnamefont {Bharti}\ \bibnamefont
  {Mahendru}}, \bibinfo {author} {\bibfnamefont {Idan}\ \bibnamefont {Tamir}},
  \bibinfo {author} {\bibfnamefont {Nils}\ \bibnamefont {Bogdanoff}}, \bibinfo
  {author} {\bibfnamefont {Olof}\ \bibnamefont {Peters}}, \bibinfo {author}
  {\bibfnamefont {Ga{\"e}l}\ \bibnamefont {Reecht}}, \bibinfo {author}
  {\bibfnamefont {Clemens~B.}\ \bibnamefont {Winkelmann}}, \bibinfo {author}
  {\bibfnamefont {Felix}\ \bibnamefont {von Oppen}}, \ and\ \bibinfo {author}
  {\bibfnamefont {Katharina~J.}\ \bibnamefont {Franke}},\ }\bibfield  {title}
  {\enquote {\bibinfo {title} {Diode effect in josephson junctions with a
  single magnetic atom},}\ }\href {\doibase 10.1038/s41586-023-05743-z}
  {\bibfield  {journal} {\bibinfo  {journal} {Nature}\ }\textbf {\bibinfo
  {volume} {615}},\ \bibinfo {pages} {628--633} (\bibinfo {year}
  {2023})}\BibitemShut {NoStop}%
\bibitem [{\citenamefont {Steiner}\ \emph {et~al.}(2023)\citenamefont
  {Steiner}, \citenamefont {Melischek}, \citenamefont {Trahms}, \citenamefont
  {Franke},\ and\ \citenamefont {von Oppen}}]{SteinerVonOppen23}%
  \BibitemOpen
  \bibfield  {author} {\bibinfo {author} {\bibfnamefont {Jacob~F.}\
  \bibnamefont {Steiner}}, \bibinfo {author} {\bibfnamefont {Larissa}\
  \bibnamefont {Melischek}}, \bibinfo {author} {\bibfnamefont {Martina}\
  \bibnamefont {Trahms}}, \bibinfo {author} {\bibfnamefont {Katharina~J.}\
  \bibnamefont {Franke}}, \ and\ \bibinfo {author} {\bibfnamefont {Felix}\
  \bibnamefont {von Oppen}},\ }\bibfield  {title} {\enquote {\bibinfo {title}
  {Diode effects in current-biased josephson junctions},}\ }\href {\doibase
  10.1103/PhysRevLett.130.177002} {\bibfield  {journal} {\bibinfo  {journal}
  {Phys. Rev. Lett.}\ }\textbf {\bibinfo {volume} {130}},\ \bibinfo {pages}
  {177002} (\bibinfo {year} {2023})}\BibitemShut {NoStop}%
\bibitem [{\citenamefont {Volkov}(1995)}]{Volkov95}%
  \BibitemOpen
  \bibfield  {author} {\bibinfo {author} {\bibfnamefont {A.~F.}\ \bibnamefont
  {Volkov}},\ }\bibfield  {title} {\enquote {\bibinfo {title} {New phenomena in
  {J}osephson {SINIS} junctions},}\ }\href {\doibase
  10.1103/PhysRevLett.74.4730} {\bibfield  {journal} {\bibinfo  {journal}
  {Phys. Rev. Lett.}\ }\textbf {\bibinfo {volume} {74}},\ \bibinfo {pages}
  {4730--4733} (\bibinfo {year} {1995})}\BibitemShut {NoStop}%
\bibitem [{\citenamefont {Golubov}\ \emph {et~al.}(1997)\citenamefont
  {Golubov}, \citenamefont {Wilhelm},\ and\ \citenamefont
  {Zaikin}}]{GolubovWilhelmZaikin97}%
  \BibitemOpen
  \bibfield  {author} {\bibinfo {author} {\bibfnamefont {A.~A.}\ \bibnamefont
  {Golubov}}, \bibinfo {author} {\bibfnamefont {F.~K.}\ \bibnamefont
  {Wilhelm}}, \ and\ \bibinfo {author} {\bibfnamefont {A.~D.}\ \bibnamefont
  {Zaikin}},\ }\bibfield  {title} {\enquote {\bibinfo {title} {Coherent charge
  transport in metallic proximity structures},}\ }\href {\doibase
  10.1103/PhysRevB.55.1123} {\bibfield  {journal} {\bibinfo  {journal}
  {Physical Review B}\ }\textbf {\bibinfo {volume} {55}},\ \bibinfo {pages}
  {1123--1137} (\bibinfo {year} {1997})},\ \bibinfo {note} {publisher: American
  Physical Society}\BibitemShut {NoStop}%
\bibitem [{\citenamefont {Wilhelm}\ \emph {et~al.}(1998)\citenamefont
  {Wilhelm}, \citenamefont {Sch\"on},\ and\ \citenamefont
  {Zaikin}}]{WilhelmZaikin98}%
  \BibitemOpen
  \bibfield  {author} {\bibinfo {author} {\bibfnamefont {Frank~K.}\
  \bibnamefont {Wilhelm}}, \bibinfo {author} {\bibfnamefont {Gerd}\
  \bibnamefont {Sch\"on}}, \ and\ \bibinfo {author} {\bibfnamefont {Andrei~D.}\
  \bibnamefont {Zaikin}},\ }\bibfield  {title} {\enquote {\bibinfo {title}
  {Mesoscopic superconducting--normal metal--superconducting transistor},}\
  }\href {\doibase 10.1103/PhysRevLett.81.1682} {\bibfield  {journal} {\bibinfo
   {journal} {Phys. Rev. Lett.}\ }\textbf {\bibinfo {volume} {81}},\ \bibinfo
  {pages} {1682--1685} (\bibinfo {year} {1998})}\BibitemShut {NoStop}%
\bibitem [{\citenamefont {Virtanen}\ and\ \citenamefont
  {Heikkil\"a}(2004)}]{VirtanenHeikkila04}%
  \BibitemOpen
  \bibfield  {author} {\bibinfo {author} {\bibfnamefont {Pauli}\ \bibnamefont
  {Virtanen}}\ and\ \bibinfo {author} {\bibfnamefont {Tero~T.}\ \bibnamefont
  {Heikkil\"a}},\ }\bibfield  {title} {\enquote {\bibinfo {title} {Thermopower
  induced by a supercurrent in superconductor--normal-metal structures},}\
  }\href {\doibase 10.1103/PhysRevLett.92.177004} {\bibfield  {journal}
  {\bibinfo  {journal} {Phys. Rev. Lett.}\ }\textbf {\bibinfo {volume} {92}},\
  \bibinfo {pages} {177004} (\bibinfo {year} {2004})}\BibitemShut {NoStop}%
\bibitem [{\citenamefont {Titov}(2008)}]{Titov08}%
  \BibitemOpen
  \bibfield  {author} {\bibinfo {author} {\bibfnamefont {M.}~\bibnamefont
  {Titov}},\ }\bibfield  {title} {\enquote {\bibinfo {title} {Thermopower
  oscillations in mesoscopic {Andreev} interferometers},}\ }\href {\doibase
  10.1103/PhysRevB.78.224521} {\bibfield  {journal} {\bibinfo  {journal}
  {Physical Review B}\ }\textbf {\bibinfo {volume} {78}},\ \bibinfo {pages}
  {224521} (\bibinfo {year} {2008})},\ \bibinfo {note} {publisher: American
  Physical Society}\BibitemShut {NoStop}%
\bibitem [{\citenamefont {Dolgirev}\ \emph {et~al.}(2018)\citenamefont
  {Dolgirev}, \citenamefont {Kalenkov},\ and\ \citenamefont
  {Zaikin}}]{DolgirevKalenkovZaikin18}%
  \BibitemOpen
  \bibfield  {author} {\bibinfo {author} {\bibfnamefont {Pavel~E.}\
  \bibnamefont {Dolgirev}}, \bibinfo {author} {\bibfnamefont {Mikhail~S.}\
  \bibnamefont {Kalenkov}}, \ and\ \bibinfo {author} {\bibfnamefont
  {Andrei~D.}\ \bibnamefont {Zaikin}},\ }\bibfield  {title} {\enquote {\bibinfo
  {title} {Current-phase relation and flux-dependent thermoelectricity in
  {Andreev} interferometers},}\ }\href {\doibase 10.1103/PhysRevB.97.054521}
  {\bibfield  {journal} {\bibinfo  {journal} {Phys. Rev. B}\ }\textbf {\bibinfo
  {volume} {97}},\ \bibinfo {pages} {054521} (\bibinfo {year}
  {2018})}\BibitemShut {NoStop}%
\bibitem [{\citenamefont {Kalenkov}\ and\ \citenamefont
  {Zaikin}(2021)}]{KalenkovZaikin21}%
  \BibitemOpen
  \bibfield  {author} {\bibinfo {author} {\bibfnamefont {M.~S.}\ \bibnamefont
  {Kalenkov}}\ and\ \bibinfo {author} {\bibfnamefont {A.~D.}\ \bibnamefont
  {Zaikin}},\ }\bibfield  {title} {\enquote {\bibinfo {title} {Phase-{Coherent}
  {Thermoelectricity} in {Superconducting} {Hybrids} ({Brief} {Review})},}\
  }\href {\doibase 10.1134/S0021364021220021} {\bibfield  {journal} {\bibinfo
  {journal} {JETP Letters}\ }\textbf {\bibinfo {volume} {114}},\ \bibinfo
  {pages} {593--608} (\bibinfo {year} {2021})}\BibitemShut {NoStop}%
\bibitem [{\citenamefont {Baselmans}\ \emph {et~al.}(1999)\citenamefont
  {Baselmans}, \citenamefont {Morpurgo}, \citenamefont {van Wees},\ and\
  \citenamefont {Klapwijk}}]{Baselmans99}%
  \BibitemOpen
  \bibfield  {author} {\bibinfo {author} {\bibfnamefont {J.~J.~A.}\
  \bibnamefont {Baselmans}}, \bibinfo {author} {\bibfnamefont {A.~F.}\
  \bibnamefont {Morpurgo}}, \bibinfo {author} {\bibfnamefont {B.~J.}\
  \bibnamefont {van Wees}}, \ and\ \bibinfo {author} {\bibfnamefont {T.~M.}\
  \bibnamefont {Klapwijk}},\ }\bibfield  {title} {\enquote {\bibinfo {title}
  {Reversing the direction of the supercurrent in a controllable {Josephson}
  junction},}\ }\href {\doibase 10.1038/16204} {\bibfield  {journal} {\bibinfo
  {journal} {Nature}\ }\textbf {\bibinfo {volume} {397}},\ \bibinfo {pages}
  {43--45} (\bibinfo {year} {1999})},\ \bibinfo {note} {publisher: Nature
  Publishing Group}\BibitemShut {NoStop}%
\bibitem [{\citenamefont {Shaikhaidarov}\ \emph {et~al.}(2000)\citenamefont
  {Shaikhaidarov}, \citenamefont {Volkov}, \citenamefont {Takayanagi},
  \citenamefont {Petrashov},\ and\ \citenamefont
  {Delsing}}]{ShaikhaidarovVolkov00}%
  \BibitemOpen
  \bibfield  {author} {\bibinfo {author} {\bibfnamefont {R.}~\bibnamefont
  {Shaikhaidarov}}, \bibinfo {author} {\bibfnamefont {A.~F.}\ \bibnamefont
  {Volkov}}, \bibinfo {author} {\bibfnamefont {H.}~\bibnamefont {Takayanagi}},
  \bibinfo {author} {\bibfnamefont {V.~T.}\ \bibnamefont {Petrashov}}, \ and\
  \bibinfo {author} {\bibfnamefont {P.}~\bibnamefont {Delsing}},\ }\bibfield
  {title} {\enquote {\bibinfo {title} {Josephson effects in a
  superconductor--normal-metal mesoscopic structure with a dangling
  superconducting arm},}\ }\href {\doibase 10.1103/PhysRevB.62.R14649}
  {\bibfield  {journal} {\bibinfo  {journal} {Phys. Rev. B}\ }\textbf {\bibinfo
  {volume} {62}},\ \bibinfo {pages} {R14649--R14652} (\bibinfo {year}
  {2000})}\BibitemShut {NoStop}%
\bibitem [{\citenamefont {Bezuglyi}\ \emph {et~al.}(2003)\citenamefont
  {Bezuglyi}, \citenamefont {Shumeiko},\ and\ \citenamefont
  {Wendin}}]{Bezuglyi03}%
  \BibitemOpen
  \bibfield  {author} {\bibinfo {author} {\bibfnamefont {E.~V.}\ \bibnamefont
  {Bezuglyi}}, \bibinfo {author} {\bibfnamefont {V.~S.}\ \bibnamefont
  {Shumeiko}}, \ and\ \bibinfo {author} {\bibfnamefont {G.}~\bibnamefont
  {Wendin}},\ }\bibfield  {title} {\enquote {\bibinfo {title} {Nonequilibrium
  {Josephson} effect in short-arm diffusive {SNS} interferometers},}\ }\href
  {\doibase 10.1103/PhysRevB.68.134506} {\bibfield  {journal} {\bibinfo
  {journal} {Physical Review B}\ }\textbf {\bibinfo {volume} {68}},\ \bibinfo
  {pages} {134506} (\bibinfo {year} {2003})},\ \bibinfo {note} {publisher:
  American Physical Society}\BibitemShut {NoStop}%
\bibitem [{\citenamefont {Sun}\ and\ \citenamefont
  {Linder}(2024)}]{SunLinder24}%
  \BibitemOpen
  \bibfield  {author} {\bibinfo {author} {\bibfnamefont {Chi}\ \bibnamefont
  {Sun}}\ and\ \bibinfo {author} {\bibfnamefont {Jacob}\ \bibnamefont
  {Linder}},\ }\bibfield  {title} {\enquote {\bibinfo {title} {Josephson
  transistor and robust supercurrent enhancement with spin-split
  superconductors},}\ }\href {\doibase 10.1103/PhysRevB.110.224512} {\bibfield
  {journal} {\bibinfo  {journal} {Physical Review B}\ }\textbf {\bibinfo
  {volume} {110}},\ \bibinfo {pages} {224512} (\bibinfo {year} {2024})},\
  \bibinfo {note} {publisher: American Physical Society}\BibitemShut {NoStop}%
\bibitem [{\citenamefont {Hijano}\ \emph {et~al.}(2021)\citenamefont {Hijano},
  \citenamefont {Ili\ifmmode~\acute{c}\else \'{c}\fi{}},\ and\ \citenamefont
  {Bergeret}}]{HijanoIlicBergeret21}%
  \BibitemOpen
  \bibfield  {author} {\bibinfo {author} {\bibfnamefont {Alberto}\ \bibnamefont
  {Hijano}}, \bibinfo {author} {\bibfnamefont {Stefan}\ \bibnamefont
  {Ili\ifmmode~\acute{c}\else \'{c}\fi{}}}, \ and\ \bibinfo {author}
  {\bibfnamefont {F.~Sebasti\'an}\ \bibnamefont {Bergeret}},\ }\bibfield
  {title} {\enquote {\bibinfo {title} {Anomalous andreev interferometer: Study
  of an anomalous josephson junction coupled to a normal wire},}\ }\href
  {\doibase 10.1103/PhysRevB.104.214515} {\bibfield  {journal} {\bibinfo
  {journal} {Phys. Rev. B}\ }\textbf {\bibinfo {volume} {104}},\ \bibinfo
  {pages} {214515} (\bibinfo {year} {2021})}\BibitemShut {NoStop}%
\bibitem [{\citenamefont {Margineda}\ \emph {et~al.}(2023)\citenamefont
  {Margineda}, \citenamefont {Claydon}, \citenamefont {Qejvanaj},\ and\
  \citenamefont {Checkley}}]{MarginedaCheckley23}%
  \BibitemOpen
  \bibfield  {author} {\bibinfo {author} {\bibfnamefont {Daniel}\ \bibnamefont
  {Margineda}}, \bibinfo {author} {\bibfnamefont {Jill~S.}\ \bibnamefont
  {Claydon}}, \bibinfo {author} {\bibfnamefont {Fatjon}\ \bibnamefont
  {Qejvanaj}}, \ and\ \bibinfo {author} {\bibfnamefont {Chris}\ \bibnamefont
  {Checkley}},\ }\bibfield  {title} {\enquote {\bibinfo {title} {Observation of
  anomalous {Josephson} effect in nonequilibrium {Andreev} interferometers},}\
  }\href {\doibase 10.1103/PhysRevB.107.L100502} {\bibfield  {journal}
  {\bibinfo  {journal} {Physical Review B}\ }\textbf {\bibinfo {volume}
  {107}},\ \bibinfo {pages} {L100502} (\bibinfo {year} {2023})},\ \bibinfo
  {note} {publisher: American Physical Society}\BibitemShut {NoStop}%
\bibitem [{\citenamefont {Daido}\ \emph {et~al.}(2025)\citenamefont {Daido},
  \citenamefont {Yanase},\ and\ \citenamefont {Law}}]{DaidoYanaseLaw25}%
  \BibitemOpen
  \bibfield  {author} {\bibinfo {author} {\bibfnamefont {Akito}\ \bibnamefont
  {Daido}}, \bibinfo {author} {\bibfnamefont {Youichi}\ \bibnamefont {Yanase}},
  \ and\ \bibinfo {author} {\bibfnamefont {Kam~Tuen}\ \bibnamefont {Law}},\
  }\href {\doibase 10.48550/arXiv.2503.16923} {\enquote {\bibinfo {title}
  {Nonreciprocal current-induced zero-resistance state in valley-polarized
  superconductors},}\ } (\bibinfo {year} {2025}),\ \bibinfo {note}
  {arXiv:2503.16923 [cond-mat]}\BibitemShut {NoStop}%
\bibitem [{\citenamefont {Lin}\ \emph {et~al.}(2022)\citenamefont {Lin},
  \citenamefont {Siriviboon}, \citenamefont {Scammell}, \citenamefont {Liu},
  \citenamefont {Rhodes}, \citenamefont {Watanabe}, \citenamefont {Taniguchi},
  \citenamefont {Hone}, \citenamefont {Scheurer},\ and\ \citenamefont
  {Li}}]{LinScheurerLi22}%
  \BibitemOpen
  \bibfield  {author} {\bibinfo {author} {\bibfnamefont {Jiang-Xiazi}\
  \bibnamefont {Lin}}, \bibinfo {author} {\bibfnamefont {Phum}\ \bibnamefont
  {Siriviboon}}, \bibinfo {author} {\bibfnamefont {Harley~D.}\ \bibnamefont
  {Scammell}}, \bibinfo {author} {\bibfnamefont {Song}\ \bibnamefont {Liu}},
  \bibinfo {author} {\bibfnamefont {Daniel}\ \bibnamefont {Rhodes}}, \bibinfo
  {author} {\bibfnamefont {K.}~\bibnamefont {Watanabe}}, \bibinfo {author}
  {\bibfnamefont {T.}~\bibnamefont {Taniguchi}}, \bibinfo {author}
  {\bibfnamefont {James}\ \bibnamefont {Hone}}, \bibinfo {author}
  {\bibfnamefont {Mathias~S.}\ \bibnamefont {Scheurer}}, \ and\ \bibinfo
  {author} {\bibfnamefont {J.~I.~A.}\ \bibnamefont {Li}},\ }\bibfield  {title}
  {\enquote {\bibinfo {title} {Zero-field superconducting diode effect in
  small-twist-angle trilayer graphene},}\ }\href {\doibase
  10.1038/s41567-022-01700-1} {\bibfield  {journal} {\bibinfo  {journal}
  {Nature Physics}\ ,\ \bibinfo {pages} {1--7}} (\bibinfo {year} {2022})},\
  \bibinfo {note} {publisher: Nature Publishing Group}\BibitemShut {NoStop}%
\bibitem [{\citenamefont {Kopnin}(2001)}]{Kopnin01}%
  \BibitemOpen
  \bibfield  {author} {\bibinfo {author} {\bibfnamefont {Nikolai~B}\
  \bibnamefont {Kopnin}},\ }\href@noop {} {\emph {\bibinfo {title} {Theory of
  nonequilibrium superconductivity}}},\ Vol.\ \bibinfo {volume} {110}\
  (\bibinfo  {publisher} {Oxford University Press},\ \bibinfo {year}
  {2001})\BibitemShut {NoStop}%
\bibitem [{\citenamefont {Golubov}\ \emph {et~al.}(2004)\citenamefont
  {Golubov}, \citenamefont {Kupriyanov},\ and\ \citenamefont
  {Il'ichev}}]{GolubovKupriyanov04}%
  \BibitemOpen
  \bibfield  {author} {\bibinfo {author} {\bibfnamefont {A.~A.}\ \bibnamefont
  {Golubov}}, \bibinfo {author} {\bibfnamefont {M.~Yu.}\ \bibnamefont
  {Kupriyanov}}, \ and\ \bibinfo {author} {\bibfnamefont {E.}~\bibnamefont
  {Il'ichev}},\ }\bibfield  {title} {\enquote {\bibinfo {title} {The
  current-phase relation in josephson junctions},}\ }\href {\doibase
  10.1103/RevModPhys.76.411} {\bibfield  {journal} {\bibinfo  {journal} {Rev.
  Mod. Phys.}\ }\textbf {\bibinfo {volume} {76}},\ \bibinfo {pages} {411--469}
  (\bibinfo {year} {2004})}\BibitemShut {NoStop}%
\bibitem [{Note*()}]{Note*}%
  \BibitemOpen
  \bibinfo {note} {From a phenomenological Ginzburg-Landau perspective, the
  term \(J_{s,0}^{(neq)}\) can be interpreted as arising from a contribution
  \(\protect \mathcal {F}_{s,0} = J_{s,0}^{(neq)} \varphi \) to the free
  energy, analogous to the \(\protect \mathbf {J} \cdot \protect \mathbf {A}\)
  coupling via gauge invariance. This interpretation also explains why no JDE
  arises solely from a thermal gradient: a heat current \(J_q\) does not couple
  to the vector potential or, consequently, to the phase difference \(\varphi
  \). Similar constant terms in the CPR have also been noted to arise due to
  screening and persistent supercurrents \cite
  {CuozzoShabani24,Cuozzo25}.}\BibitemShut {Stop}%
\bibitem [{\citenamefont {Amin}\ \emph {et~al.}(2001)\citenamefont {Amin},
  \citenamefont {Omelyanchouk},\ and\ \citenamefont {Zagoskin}}]{Amin01}%
  \BibitemOpen
  \bibfield  {author} {\bibinfo {author} {\bibfnamefont {M.~H.~S.}\
  \bibnamefont {Amin}}, \bibinfo {author} {\bibfnamefont {A.~N.}\ \bibnamefont
  {Omelyanchouk}}, \ and\ \bibinfo {author} {\bibfnamefont {A.~M.}\
  \bibnamefont {Zagoskin}},\ }\bibfield  {title} {\enquote {\bibinfo {title}
  {Mesoscopic multiterminal {Josephson} structures. {I}. {Effects} of nonlocal
  weak coupling},}\ }\href {\doibase 10.1063/1.1399198} {\bibfield  {journal}
  {\bibinfo  {journal} {Low Temperature Physics}\ }\textbf {\bibinfo {volume}
  {27}},\ \bibinfo {pages} {616--623} (\bibinfo {year} {2001})}\BibitemShut
  {NoStop}%
\bibitem [{\citenamefont {Pankratova}\ \emph {et~al.}(2020)\citenamefont
  {Pankratova}, \citenamefont {Lee}, \citenamefont {Kuzmin}, \citenamefont
  {Wickramasinghe}, \citenamefont {Mayer}, \citenamefont {Yuan}, \citenamefont
  {Vavilov}, \citenamefont {Shabani},\ and\ \citenamefont
  {Manucharyan}}]{PankratovaShabani20}%
  \BibitemOpen
  \bibfield  {author} {\bibinfo {author} {\bibfnamefont {Natalia}\ \bibnamefont
  {Pankratova}}, \bibinfo {author} {\bibfnamefont {Hanho}\ \bibnamefont {Lee}},
  \bibinfo {author} {\bibfnamefont {Roman}\ \bibnamefont {Kuzmin}}, \bibinfo
  {author} {\bibfnamefont {Kaushini}\ \bibnamefont {Wickramasinghe}}, \bibinfo
  {author} {\bibfnamefont {William}\ \bibnamefont {Mayer}}, \bibinfo {author}
  {\bibfnamefont {Joseph}\ \bibnamefont {Yuan}}, \bibinfo {author}
  {\bibfnamefont {Maxim~G.}\ \bibnamefont {Vavilov}}, \bibinfo {author}
  {\bibfnamefont {Javad}\ \bibnamefont {Shabani}}, \ and\ \bibinfo {author}
  {\bibfnamefont {Vladimir~E.}\ \bibnamefont {Manucharyan}},\ }\bibfield
  {title} {\enquote {\bibinfo {title} {Multiterminal josephson effect},}\
  }\href {\doibase 10.1103/PhysRevX.10.031051} {\bibfield  {journal} {\bibinfo
  {journal} {Phys. Rev. X}\ }\textbf {\bibinfo {volume} {10}},\ \bibinfo
  {pages} {031051} (\bibinfo {year} {2020})}\BibitemShut {NoStop}%
\bibitem [{\citenamefont {Osin}\ \emph {et~al.}(2024)\citenamefont {Osin},
  \citenamefont {Levchenko},\ and\ \citenamefont
  {Khodas}}]{OsinLevchenkoKhodas24}%
  \BibitemOpen
  \bibfield  {author} {\bibinfo {author} {\bibfnamefont {A.~S.}\ \bibnamefont
  {Osin}}, \bibinfo {author} {\bibfnamefont {Alex}\ \bibnamefont {Levchenko}},
  \ and\ \bibinfo {author} {\bibfnamefont {Maxim}\ \bibnamefont {Khodas}},\
  }\bibfield  {title} {\enquote {\bibinfo {title} {Anomalous {Josephson} diode
  effect in superconducting multilayers},}\ }\href {\doibase
  10.1103/PhysRevB.109.184512} {\bibfield  {journal} {\bibinfo  {journal}
  {Physical Review B}\ }\textbf {\bibinfo {volume} {109}},\ \bibinfo {pages}
  {184512} (\bibinfo {year} {2024})},\ \bibinfo {note} {publisher: American
  Physical Society}\BibitemShut {NoStop}%
\bibitem [{\citenamefont {V\'{a}vra}\ \emph {et~al.}(2013)\citenamefont
  {V\'{a}vra}, \citenamefont {Pfaff}, \citenamefont {Monaco}, \citenamefont
  {Aprili},\ and\ \citenamefont {Strunk}}]{VarvaStrunk13}%
  \BibitemOpen
  \bibfield  {author} {\bibinfo {author} {\bibfnamefont {O.}~\bibnamefont
  {V\'{a}vra}}, \bibinfo {author} {\bibfnamefont {W.}~\bibnamefont {Pfaff}},
  \bibinfo {author} {\bibfnamefont {R.}~\bibnamefont {Monaco}}, \bibinfo
  {author} {\bibfnamefont {M.}~\bibnamefont {Aprili}}, \ and\ \bibinfo {author}
  {\bibfnamefont {C.}~\bibnamefont {Strunk}},\ }\bibfield  {title} {\enquote
  {\bibinfo {title} {Current-controllable planar {S-(S/F)-S} {J}osephson
  junction},}\ }\href {\doibase 10.1063/1.4792213} {\bibfield  {journal}
  {\bibinfo  {journal} {Applied Physics Letters}\ }\textbf {\bibinfo {volume}
  {102}},\ \bibinfo {pages} {072602} (\bibinfo {year} {2013})}\BibitemShut
  {NoStop}%
\bibitem [{\citenamefont {Salamone}\ \emph {et~al.}(2021)\citenamefont
  {Salamone}, \citenamefont {Svendsen}, \citenamefont {Amundsen},\ and\
  \citenamefont {Jacobsen}}]{Salamone21}%
  \BibitemOpen
  \bibfield  {author} {\bibinfo {author} {\bibfnamefont {Tancredi}\
  \bibnamefont {Salamone}}, \bibinfo {author} {\bibfnamefont {Mathias B.~M.}\
  \bibnamefont {Svendsen}}, \bibinfo {author} {\bibfnamefont {Morten}\
  \bibnamefont {Amundsen}}, \ and\ \bibinfo {author} {\bibfnamefont {Sol}\
  \bibnamefont {Jacobsen}},\ }\bibfield  {title} {\enquote {\bibinfo {title}
  {Curvature-induced long-range supercurrents in diffusive
  superconductor-ferromagnet-superconductor {J}osephson junctions with a
  dynamic $0-\pi$ transition},}\ }\href {\doibase 10.1103/PhysRevB.104.L060505}
  {\bibfield  {journal} {\bibinfo  {journal} {Phys. Rev. B}\ }\textbf {\bibinfo
  {volume} {104}},\ \bibinfo {pages} {L060505} (\bibinfo {year}
  {2021})}\BibitemShut {NoStop}%
\bibitem [{\citenamefont {Salamone}\ \emph {et~al.}(2022)\citenamefont
  {Salamone}, \citenamefont {Hugdal}, \citenamefont {Amundsen},\ and\
  \citenamefont {Jacobsen}}]{Salamone22}%
  \BibitemOpen
  \bibfield  {author} {\bibinfo {author} {\bibfnamefont {Tancredi}\
  \bibnamefont {Salamone}}, \bibinfo {author} {\bibfnamefont {Henning~G.}\
  \bibnamefont {Hugdal}}, \bibinfo {author} {\bibfnamefont {Morten}\
  \bibnamefont {Amundsen}}, \ and\ \bibinfo {author} {\bibfnamefont {Sol~H.}\
  \bibnamefont {Jacobsen}},\ }\bibfield  {title} {\enquote {\bibinfo {title}
  {Curvature control of the superconducting proximity effect in diffusive
  ferromagnetic nanowires},}\ }\href {\doibase 10.1103/PhysRevB.105.134511}
  {\bibfield  {journal} {\bibinfo  {journal} {Phys. Rev. B}\ }\textbf {\bibinfo
  {volume} {105}},\ \bibinfo {pages} {134511} (\bibinfo {year}
  {2022})}\BibitemShut {NoStop}%
\end{thebibliography}%

\end{document}